\documentstyle[12pt]{article}                  
\input{amssymb.sty}
\makeatletter
\def\nothing#1{}
\newdimen\earraycolsep
\renewcommand{\thetable}{\arabic{table}}
\renewcommand{\thefigure}{\arabic{figure}}
\renewcommand{\title}[1]{%
 \vspace*{120\p@}%
  {\parindent \z@ \raggedright \reset@font
    \bfseries #1\par
    \nobreak
    \vskip 36\p@
  }}
\def\author#1{{\pretolerance=10000 \raggedright \advance \leftskip by 1in
\noindent #1 \vskip 1pc}}
\def\affiliation#1{{\advance\leftskip by 1in \noindent #1 \vskip -1pc}}
\def\refnote#1{{$^{\hbox{\scriptsize #1}}$}}
\def\affnote#1{\llap{$^{\hbox{\scriptsize #1}}$}}

\renewcommand\section{\@startsection{section}{1}{\z@}{2pc \@plus
      1ex minus .2ex}{1pc \@plus .2ex}{\reset@font
      \normalsize\bfseries\noindent
      {\addtocounter{section}{1}}\Roman{section}\
      {\setcounter{subsection}{0}
      \setcounter{subsubsection}{0}\setcounter{equation}{0}} }}
\renewcommand\subsection{\@startsection{subsection}{2}{\z@}{1pc \@plus 1ex
    minus.2ex}{1pc \@plus .2ex}
    {\reset@font\normalsize\bfseries
    \noindent{\addtocounter{subsection}{1}}%
    {\setcounter{subsubsection}{0}}\Roman{section}.\Roman{subsection}\ }}
\renewcommand\subsubsection{\@startsection{subsubsection}{3}{\parindent}
        {1pc \@plus 1ex minus.2ex}{-0.5em}{\reset@font\normalsize\bfseries%
        {\addtocounter{subsubsection}{1}} \hspace*{.6cm}
        \Roman{section}.\Roman{subsection}.\Roman{subsubsection}
        \hspace*{-7mm}}}

\def\AmS{{\protect\the\textfont2%
        A\kern-.1667em\lower.5ex\hbox{M}\kern-.125emS}}

\def\p@LaTeX{{\family{times}\series{m}\shape{n}\selectfont
L\kern-.36em\raise.3ex\hbox{\scriptsize A}\kern-.15em
T\kern-.1667em\lower.7ex\hbox{E}\kern-.125emX}}

\newlength{\colwidth}

\def\@oddhead{\hfil}
\def\@evenhead{\hfil}
\def\@oddfoot{{\bfseries\hfil\thepage}}
\def\@evenfoot{{\bfseries\thepage\hfil}}
\def\fnum@figure{\footnotesize\raggedright{\bfseries \figurename~\thefigure.}}
\def\fnum@table{\normalsize\raggedright{\bfseries \tablename~\thetable.}}
\long\def\@makecaption#1#2{\vskip 10\p@ {#1 #2\par}}
\long\def\@makefntext#1{\setbox0=\hbox{$\m@th^{\@thefnmark}$}\noindent
\hangindent=\wd0 \box0 #1}
\newbox\@atbox
\long\def\atable#1#2#3{\begin{table}[tbp]\centering\footnotesize
\setbox\@atbox\hbox{#2}
\parbox{\wd\@atbox}{\caption{#1}}\par\smallskip
#2
\par\smallskip\parbox{\wd\@atbox}{\raggedright #3}
\end{table}}

\font\goth=eufm10

\def\Cb{{\mathbb C}}
\def\Hb{{\mathbb H}}
\def\Nb{{\mathbb N}}
\def\Rb{{\mathbb R}}
\def\Tb{{\mathbb T}}
\def\Zb{{\mathbb Z}}

\def\Ac{{\cal A}}
\def\Bc{{\cal B}}

\def\Ec{{\cal E}}
\def\Hc{{\cal H}}

\def\Kc{{\cal K}}
\def\Lc{{\cal L}}

\def\Sc{{\cal S}}
\def\Uc{{\cal U}}
\newcommand{\wt}{\widetilde}

\def\a{\alpha}
\def\b{\beta}
\def\d{\delta}
\def\lb{\lambda}
\def\g{\gamma}
\def\om{\omega}
\def\s{\sigma}
\def\t{\theta}
\def\ve{\varepsilon}
\def\vp{\varphi}

\def\z{\zeta}

\def\D{\Delta}
\def\G{\Gamma}
\def\Lb{\Lambda}
\def\Om{\Omega}
\def\Si{\Sigma}

\def\ex{\exists}
\def\fl{\forall}
\def\ify{\infty}
\def\lgl{\langle}
\def\nb{\nabla}
\def\op{\oplus}
\def\ot{\otimes}
\def\ov{\overline}
\def\part{\partial}
\def\rgl{\rangle}
\def\sbs{\subset}
\def\semi{>\!\!\!\lhd}
\def\sm{\simeq}
\def\ts{\times}
\def\wdg{\wedge}

\def\ra{\rightarrow}
\def\longra{\longrightarrow}
\def\Ra{\Rightarrow}
\def\lra{\leftrightarrow}
\def\text{\hbox}

\def\Aut{\mathop{\rm Aut}\nolimits}
\def\Diff{\mathop{\rm Diff}\nolimits}
\def\Dom{\mathop{\rm Dom}\nolimits}
\def\End{\mathop{\rm End}\nolimits}
\def\id{\mathop{\rm id}\nolimits}
\def\Index{\mathop{\rm Index}\nolimits}

\def\Int{\mathop{\rm Int}\nolimits}
\def\Ker{\mathop{\rm Ker}\nolimits}
\def\Rang{\mathop{\rm Rang}\nolimits}
\def\Re{\mathop{\rm Re}\nolimits}
\def\Res{\mathop{\rm Res}\nolimits}
\def\Sign{\mathop{\rm Sign}\nolimits}

\def\Sup{\mathop{\rm Sup}\nolimits}
\def\Trace{\mathop{\rm Trace}\nolimits}

\def\build#1_#2^#3{\mathrel{
\mathop{\kern 0pt#1}\limits_{#2}^{#3}}}

\def\@nbibitem#1{\noindent \hangindent=2pc \hangafter=1
\refstepcounter{enumi}\hbox to 2pc{\arabic{enumi}.\hfil}%
\immediate\write\@auxout{\string\bibcite{#1}{\arabic{enumi}}}}
\def\numbibliography{%
\section*{REFERENCES}%
\bgroup\footnotesize
\setcounter{enumi}{0}%
\def\newblock{\hskip .11em plus.33em minus.07em}%
\let\bibitem\@nbibitem}
\def\endnumbibliography{\par\egroup}
\makeatother

\begin{document}

\title{\centerline{A SHORT SURVEY OF NONCOMMUTATIVE GEOMETRY}}

\author{\bf Alain CONNES,\refnote{1}}

\affiliation{\affnote{1}  Coll\`ege de France,
3, rue Ulm,
75005 PARIS\\
and\\
I.H.E.S.,
35, route de Chartres,
91440 BURES-sur-YVETTE
}

\vspace{1cm}

\begin{abstract}
We give a survey of selected topics in noncommutative geometry,  with some 
emphasis
on those directly related to physics, including our recent work with Dirk Kreimer 
on 
renormalization and the Riemann-Hilbert problem.
We discuss at length two issues.
The first is the relevance of the paradigm of geometric space, based on spectral 
considerations,
which is central in the theory. As a simple illustration of the spectral formulation
of geometry in the ordinary commutative case, we give a polynomial equation for geometries on 
the four dimensional sphere with fixed volume.
The equation involves an idempotent e, playing the role of the instanton, and 
the 
Dirac operator D. It expresses the gamma five matrix as the pairing between the operator theoretic chern characters of 
 e and D.
It is of degree five in the idempotent and four in the Dirac operator which only appears 
through its commutant with the idempotent. It 
determines both the sphere and all its metrics with
fixed volume form. 

\noindent We also show using the noncommutative analogue of the Polyakov action, 
how 
to obtain the noncommutative metric
(in spectral form) on the noncommutative tori from the formal naive metric. We 
conclude on some questions
 related to string theory. 
\end{abstract}

\vspace{1cm}

\section{Introduction}
The origin of noncommutative geometry is twofold. 

\noindent On the one hand there is a wealth of examples of 
spaces whose coordinate algebra is no longer commutative 
but which have obvious relevance in physics or mathematics. The first examples
 came from phase space in quantum mechanics but there 
are many others, such as the leaf spaces of foliations, the 
duals of nonabelian discrete groups, the space of Penrose
tilings, the Brillouin zone in solid state physics, the noncommutative tori which 
appear
naturally in
M-theory compactification, and the Adele class space
which is a natural geometric space carrying an action of 
the analogue of the Frobenius for global fields of zero characteristic. Finally 
various models
 of space-time itself are interesting examples of noncommutative spaces.

\noindent On the other hand the stretching of geometric thinking 
imposed by passing to noncommutative spaces forces one
to rethink about most of our familiar notions.
The difficulty is not to add arbitrarily the adjective
quantum to our geometric words but to
develop far reaching extensions of classical concepts, ranging from the simplest 
which is measure theory, 
to the most sophisticated which is geometry itself.

\bigskip
\section{Measure theory}

\noindent The extension of the classical concepts has been achieved a long time 
ago 
by operator
algebraists as far as measure theory is concerned.
The theory of nonabelian von Neumann algebras is
indeed a far reaching extension of measure theory,
whose main surprise is that such an algebra $M$ inherits
from its noncommutativity a god-given time evolution.

\noindent It is given by the group homomorphism, (\cite{Co_2})
\begin{equation}
\delta : \Rb \rightarrow {\rm Out} (M) = {\rm Aut} (M) / {\rm Int} (M) \, 
\label{eq:(25)}
\end{equation}
from the additive group $\Rb$ to the group of automorphism classes of $M$ modulo 
inner 
automorphisms.

\noindent This uniqueness of the, a priori state dependent, modular automorphism
 group of a state, together with the earlier work of Powers, Araki-Woods and 
Krieger were the first steps which eventually led to the complete classification 
of 
approximately finite 
dimensional factors (also called hyperfinite).

\noindent They are classified by their module,
\begin{equation}
{\rm Mod} (M) \mathop{\subset}_{\sim} \ \Rb_+^* \, ,  \label{eq:(20)}
\end{equation}
which is a virtual closed subgroup of $\Rb_+^*$ in the sense of G.~Mackey, i.e. an 
ergodic
action of $\Rb_+^*$.  

\medskip

The classification involves three independent parts,
\begin{itemize}
\item[(A)] The definition of the invariant ${\rm Mod} (M)$ for arbitrary  
factors.
\item[(B)] The equivalence of all possible notions of approximate finite  
dimensionality.
\item[(C)] The proof that Mod is a complete invariant and that all 
virtual subgroups are obtained.
\end{itemize}
The module of a factor $M$ was first defined (\cite{Co_2}) as a closed 
subgroup of $\Rb_+^*$ by the equality
\begin{equation}
S(M) = \bigcap_{\varphi} \ {\rm Spec} (\Delta_{\varphi}) \subset \Rb_+ ,  
\label{eq:(21)}
\end{equation}
where $\varphi$ varies among (faithful, normal) states on $M$ and the operator 
$\Delta_{\varphi}$ is the {\it modular operator} of the Tomita-Takesaki theory
(\cite{T}).

\noindent The virtual subgroup ${\rm Mod} (M)$ 
is the {\it flow of weights} ( \cite{Co_2} \cite{Ta} \cite{K} \cite{CT}) of $M$. 
It 
is 
obtained from the module $\delta$ as the dual action of $\Rb_+^*$ on the abelian 
algebra,
\begin{equation}
C = \hbox{Center of} \,( \ M \, {\textstyle \semi_{\delta}} \ \Rb )\, ,  
\label{eq:(28)}
\end{equation} 
where $M \, \semi_{\delta} \ \Rb$ is the crossed product of $M$ by the 
modular automorphism group $\delta$.

\noindent This takes care of (A), to describe (B) let us simply state the 
equivalence (\cite{Co_1}) of the following conditions
\begin{equation}
\matrix{
&M \ \hbox{is the closure of the union of an increasing sequence of} \cr  
&\hbox{finite dimensional algebras.} \hfill \cr
}   \label{eq:(29)}
\end{equation} 
\begin{equation}
\matrix{
&M \ \hbox{is complemented as a subspace of the normed space of} \cr 
&\hbox{all operators in a Hilbert space.} \hfill \cr
}   \label{eq:(30)}
\end{equation}
 
\noindent The condition (\ref{eq:(29)}) is obviously what one would expect for an 
approximately finite dimensional algebra. Condition (\ref{eq:(30)}) is similar to 
{\it amenability} for discrete groups and the implication (\ref{eq:(30)}) 
$\Rightarrow$ (\ref{eq:(29)}) is a very powerful tool.

\noindent Besides the reduction from type III to type II (\cite{Co_2} \cite{Ta}), 
the proof of (C) involves the uniqueness of the approximately finite dimensional 
factor of type ${\rm II}_{\infty}$, \cite{Co_1},
the classification of its automorphisms \cite{ccccc} for the ${{\rm 
III}_{\lambda}}$ 
case, and the 
results of Krieger  \cite{K} for the ${\rm III}_0$ case. The only case which was 
left open 
in 1976 was the ${\rm III}_1$ case, which was reduced to a problem on the 
bicentralizer of states 
\cite{Co$_{21}$}, this problem was finally settled by U. Haagerup in \cite{Ha}. 
Since then, the subject 
of von-Neumann algebras has undergone two major revolutions, thanks first to the 
famous work 
of Vaughan Jones on subfactors and then to the pioneering work of Dan Voiculescu
who created and developped the completely new field of free probability theory.

\noindent Von Neumann algebras arise very naturally in geometry 
from foliated manifolds $(V,F)$. The 
von Neumann algebra $L^{\infty} (V,F)$ of a foliated manifold is easy to describe, 
its 
elements are random operators $T = 
(T_f)$, i.e. bounded measurable families of operators $T_f$ parametrized  
by the leaves $f$ of the foliation. For each leaf $f$ the operator $T_f$ acts in 
the Hilbert space $L^2 (f)$ of square integrable densities on the 
manifold $f$. Two random operators are identified if they are equal for 
almost all leaves $f$ (i.e. a set of leaves whose union in $V$ is 
negligible). The algebraic operations of sum and product are given by,
\begin{equation}
(T_1 + T_2)_f = (T_1)_f + (T_2)_f \, , \ (T_1 \, T_2)_f = (T_1)_f \, (T_2)_f 
\,  ,  \label{eq:(38)}
\end{equation} 
i.e. are effected pointwise.

\noindent All types of factors occur from this geometric construction and the 
continuous 
dimensions of Murray and von-Neumann play an essential role in the longitudinal 
index 
theorem. 

\noindent Finally we refer to \cite{ccc} for the role of approximately finite 
dimensional factors in number theory as the missing Brauer theory at Archimedean 
places.
\bigskip
\section{Topology}

\noindent The development of the topological ideas was prompted by the work of 
Israel 
Gel'fand, whose C* algebras give the required framework for noncommutative 
topology. 
The two 
main driving forces were the Novikov conjecture on homotopy invariance of higher 
signatures 
of ordinary manifolds as well as the Atiyah-Singer Index theorem. It has led, 
through the 
work of  Atiyah, Singer, Brown, Douglas, Fillmore, Miscenko and Kasparov  
(\cite{[AT]} \cite{Singer} \cite{B-D-F} \cite{[21]} \cite{[18]}) to the 
recognition 
that not only the 
Atiyah-Hirzebruch K-theory but more importantly the dual K-homology admit Hilbert 
space 
techniques and functional analysis as their natural framework. The cycles in the 
K-homology 
group $K_*(X)$ of a compact space X are indeed given by Fredholm representations 
of 
the C* 
algebra A of continuous functions on X. The central tool is the Kasparov bivariant 
K-theory. 
A basic example of C* algebra to which the theory applies is the group ring of a 
discrete 
group and restricting oneself to commutative algebras is an obviously undesirable 
assumption.

\noindent For a $C^*$ algebra $A$, let $K_0 (A)$, $K_1 (A)$ be its $K$ theory 
groups. 
Thus $K_0 (A)$ is the algebraic $K_0$ theory of the ring $A$ and $K_1 (A)$ 
is the algebraic $K_0$ theory of the ring $A \ot C_0 (\Rb) = C_0 (\Rb , 
A)$. If $A \ra B$ is a morphism of $C^*$ algebras, then there are induced 
homomorphisms of abelian groups $K_i (A) \ra K_i (B)$. Bott periodicity 
provides a six term $K$ theory exact sequence for each exact sequence $0 
\ra J \ra A \ra B \ra 0$ of $C^*$ algebras and excision shows that the $K$ groups 
involved in 
the exact sequence only depend on the respective $C^*$ algebras. As an exercice to 
appreciate 
the power of this abstract tool one should for instance use the six term $K$ 
theory 
exact sequence
 to give a short proof of the 
Jordan curve theorem.

\noindent Discrete groups, Lie groups, group actions and foliations give rise 
through 
their convolution algebra to a canonical $C^*$ algebra, and hence to $K$ 
theory groups. The analytical meaning of these $K$ theory groups is clear 
as a receptacle for indices of elliptic operators. However, these groups 
are difficult to compute. For instance, in the case of semi-simple Lie 
groups the free abelian group with one generator for each irreducible 
discrete series representation is contained in $K_0 \, C_r^* G$ where $C_r^* G$ 
is the reduced $C^*$ algebra of $G$. Thus an explicit determination of the 
$K$ theory in this case in particular involves an enumeration of the 
discrete series.

\noindent We introduced with P. Baum (\cite{[BC]}) a geometrically defined $K$ 
theory which 
specializes to discrete groups, Lie groups, group actions, and foliations. 
Its main features are its computability and the simplicity of its 
definition. In the case of semi-simple Lie groups it elucidates the role of 
the homogeneous space $G/K$ ($K$ the maximal compact subgroup of $G$) in 
the Atiyah-Schmid geometric construction of the discrete series \cite{[4]}. Using 
elliptic operators we constructed a natural map from our geometrically 
defined $K$ theory groups to the above analytic (i.e. $C^*$ algebra) $K$ 
theory groups. Much progress has been made in the past years to 
determine the range of validity of the isomorphism between the geometrically 
defined $K$ theory groups and the above analytic (i.e. $C^*$ algebra) $K$ 
theory groups. We refer to the three Bourbaki seminars  (\cite{BBB}, \cite{BBBB}, 
\cite{BBBBB}) for an update on this topic.
\bigskip
\section{Differential Topology}

\noindent The development of differential geometric ideas,
including de Rham homology, connections and curvature
of vector bundles, etc... took place during the eighties
thanks to cyclic cohomology which came from two different horizons  
(\cite{C2} \cite{Co$_{15}$} \cite{Co$_{17}$} \cite{Co$_{18}$} \cite{Ts$_1$}).
 This led for instance
to the proof of the Novikov conjecture for hyperbolic
groups \cite{[C-M1]}, but got many other applications. Basically,
by extending the Chern-Weil characteristic classes
to the general framework it allows for many concrete
computations of differential geometric nature on noncommutative spaces.
It also showed the depth of the relation between the above classification of 
factors 
and the 
geometry of foliations. For instance, using cyclic cohomology together with the 
following 
simple fact,
\begin{equation}
\matrix{
&\hbox{``A connected group can only act trivially on a homotopy} \cr 
&\hbox{invariant cohomology theory'',} \hfill \cr
} \label{eq:(42)}
\end{equation} 
one proves (cf. \cite{[Co1]}) that for any codimension one foliation $F$ of  
a compact manifold $V$ with non vanishing Godbillon-Vey class one has,
\begin{equation}
{\rm Mod} (M) \ \hbox{has finite covolume in} \ \Rb_+^* \, ,  \label{eq:(43)}
\end{equation} 
where $M = L^{\infty} (V,F)$ and a virtual subgroup of finite covolume is 
a flow with a finite invariant measure.

\noindent In its simplest form, cyclic cohomology is the cohomology theory 
obtained 
from 
the cochain complex of ($n+1$)-linear form on $\Ac$, $n$ arbitrary, such that
\begin{equation} 
\varphi (a^0 ,a^1 ,...,a^n) = (-1)^n \varphi (a^1 , a^2 ,...,a^0) \,\quad \forall 
a_j \in 
\Ac, \label{eq:(100)}
\end{equation} 
with coboundary operator given by 
\begin{eqnarray}
\lefteqn{( b\varphi) (a^0 ,\ldots ,a^{n+1}) =} \nonumber \\
& \sum_0^n (-1)^j \, \varphi (a^0 ,\ldots ,a^j a^{j+1}, \ldots
,a^{n+1}) + (-1)^{n+1} \, \varphi (a^{n+1} a^0 ,a^1 ,\ldots ,a^{n})
\end{eqnarray}

\noindent Its first important role is to provide invariants of $K$-theory classes 
as 
follows.
Given an n-dimensional cyclic cocycle on $\Ac$, n even, the following scalar is 
invariant 
under homotopy for projectors (idempotents) $E \in M_n (\Ac)$,
\begin{equation}
\varphi_n (E,E,...,E)
\end{equation}
where $\varphi$ has been uniquely extended to $M_n (\Ac)$ using the trace on
$M_n (\Cb)$, as in (9) below.
This defines a $\hbox{pairing} \ \langle K(\Ac) , HC(\Ac) \rangle$ between cyclic 
cohomology 
and K-theory.

\noindent When we take $\Ac = C^{\infty} (M)$ for a manifold $M$ and let
\begin{equation}
\varphi (f^0 ,f^1 ,...,f^n) = \langle C , f^0 df^1 \wedge df^2\wedge \ldots \wedge 
df^n 
\rangle \quad \forall f^j \in \Ac \label{eq:(0.29)}
\end{equation}
where $C$ is an n-dimensional closed de Rham current, the above invariant is equal 
to (up to 
normalization)
\begin{equation}
\langle C , Ch (E) \rangle \label{eq:(0.30)}
\end{equation}
where $Ch (E)$ is the Chern character of the vector bundle $E$ on $M$ whose
fiber at $x \in M$ is the range of $E(x) \in M_n (\Cb)$. In this example we
see that for any permutation of $\{ 0,1,...,n \}$ one has:
\begin{equation}
\varphi (f^{\sigma (0)} , f^{\sigma (1)} ,..., f^{\sigma (n)}) = \varepsilon
(\sigma) \varphi (f^0 , f^1 ,..., f^n) \label{eq:(0.31)}
\end{equation}
where $\varepsilon (\sigma)$ is the signature of the permutation. However
when we extend $\varphi$ to $M_n (\Ac)$ as $\varphi_n = \varphi \otimes {\rm
Tr}$,
\begin{equation}
\varphi_n (f^0 \otimes \mu^0 , f^1 \otimes \mu^1 ,\ldots, f^n \otimes \mu^n) =
\varphi (f^0 , f^1 ,\ldots,f^n) {\rm Tr} (\mu^0 \mu^1 \ldots \mu^n) 
\label{eq:(0.32)}
\end{equation}
the property (8) only survives for {\it cyclic} permutations.
This is at the origin of the name, {\it cyclic cohomology}, given to the
corresponding cohomology theory.

\noindent Both the Hochschild and Cyclic cohomologies of the algebra $\Ac = 
C^{\infty} (M)$ 
of smooth functions on a manifold $M$ were computed in \cite{Co$_{17}$}, 
\cite{Co$_{18}$}, thus showing how to 
extend the familiar differential geometric notions to the general noncommutative 
case 
according to the following dictionnary:

$$
\matrix{
\hbox{Space} &\hbox{Algebra} \cr
\cr
\hbox{Vector bundle} &\hbox{Finite projective module} \cr
\cr
\hbox{Differential form} &\hbox{(Class of) Hochschild cycle} \cr
\cr
\hbox{DeRham current} &\hbox{(Class of) Hochschild cocycle} \cr
\cr
\hbox{DeRham homology} &\hbox{Cyclic cohomology} \cr
\cr
\hbox{Chern Weil theory} &\hbox{Pairing} \ \langle K(\Ac) , HC(\Ac) \rangle
\cr
} \leqno (A)
$$

A simple example of cyclic cocycle on a nonabelian group ring is provided by
the following formula. Any {\it group cocycle} $c \in H^*
(B\Gamma) = H^* (\Gamma)$ gives rise to a cyclic
cocycle $\varphi_c$ on the algebra $\Ac = \Cb \Gamma$
\begin{equation}
\varphi_c (g_0 , g_1 , \ldots , g_n) = \left\{ \matrix{ 0 \ \hbox{if} \ g_0
\ldots g_n \not= 1 \cr c(g_1 , \ldots , g_n) \ \hbox{if} \ g_0 \ldots g_n =1
\cr} \right. \label{eq:(0.39)}
\end{equation}
where $c \in Z^n (\Gamma ,\Cb)$ is suitably normalized, and \ref{eq:(0.39)}
is extended by linearity to $\Cb \Gamma$.

\noindent Cyclic cohomology has an equivalent description by means of 
the bicomplex $(b,B)$ which is given by the following operators acting on 
multi-linear forms
on $\Ac$,
\begin{eqnarray}
\lefteqn{(b\vp) (a^0 ,\ldots ,a^{n+1}) =} \nonumber \\
& \sum_0^n (-1)^j \, \vp (a^0 ,\ldots ,a^j a^{j+1}, \ldots
,a^{n+1}) + (-1)^{n+1} \, \vp (a^{n+1} a^0 ,a^1 ,\ldots ,
a^{n})                                                  \label{eq:(3.5)}\\
\lefteqn{B = AB_0 \ , \ B_0 \, \vp (a^0 ,\ldots ,a^{n-1}) = \vp
(1,a^0 ,\ldots ,a^{n-1}) - (-1)^n \, \vp (a^0 ,\ldots
,a^{n-1} ,1)} \nonumber \\
& \qquad \qquad (A\psi) (a^0 ,\ldots ,a^{n-1}) = \sum_0^{n-1}
(-1)^{(n-1)j} \, \psi (a^j ,a^{j+1}, \ldots ,a^{j-1}) \,
.                                                \label{eq:3.6}
\end{eqnarray}

 \noindent The pairing between cyclic cohomology and K-theory is given in this 
presentation by the
following formula for the Chern character of the class of an idempotent $e$, up to 
normalization one has
\begin{equation}
 Ch_n (e) = \, (e-1/2) \ot e \ot e \ot ...  \ot e,
\end{equation}
where $e$ appears 2n times in the right hand side of the equation.

At the conceptual level, cyclic cohomology is a way to embed the nonadditive 
category of 
algebras and
algebra homomorphisms in an additive category of modules. The latter is the 
additive 
category of
$\Lambda$-modules where $\Lambda$ is the cyclic category. Cyclic cohomology is 
then 
obtained as an $Ext$
functor (\cite{C2}).

\noindent The cyclic category is a small category which can be defined by
generators and relations. It has the same objects as the small
category $\D$ of totally ordered finite sets and increasing maps which plays a key 
role in simplicial topology. Let
us recall (we shall use it later) that $\D$ has one object $[n]$ for each
integer $n$, and is generated by faces $\delta_i, [n-1] \ra [n]$ (the
injection that misses $i$), and degeneracies $\s_j,[n+1] \ra [n] $
(the surjection which identifies $j$ with $j+1$), with the relations,
\begin{equation}\label{ad}
\delta_j  \delta_i = \delta_i  \delta_{j-1}
\ \hbox{for} \ i < j  , \ \s_j  \s_i = 
\s_i  \s_{j+1} \qquad i \leq j 
\end{equation}
$$
\s_j  \delta_i = \left\{ \matrix{
\delta_i  \s_{j-1} \hfill &i < j \hfill \cr
1_n \hfill &\hbox{if} \ i=j \ \hbox{or} \ i = j+1 \cr
\delta_{i-1}  \s_j \hfill &i > j+1  . \hfill \cr
} \right.
$$
To obtain $\Lambda$ one adds for each $n$ a new morphism $\tau_n, [n]
\ra [n]$ such that,
\begin{equation}\label{ae}
\matrix{
\tau_n  \delta_i = \delta_{i-1}  
\tau_{n-1} &1 \leq i \leq n , &\tau_n  \delta_0 = 
\delta_n \hfill \cr
\cr
\tau_n  \s_i = \s_{i-1} 
\tau_{n+1} &1 \leq i \leq n , &\tau_n  \s_0 = 
\s_n  \tau_{n+1}^2 \cr
\cr
\tau_n^{n+1} = 1_n  . \hfill \cr
} 
\end{equation}

 The original definition of $\Lambda$ (cf.~\cite{C2})
used homotopy classes of non decreasing maps from $S^1$ to $S^1$ of
degree~1, mapping $\Zb / n$ to $\Zb / m$ and is trivially equivalent
to the above. 

\noindent Given an algebra $A$ one obtains a module over the small category
$\Lambda$ by assigning to each integer $n \geq 0$ the vector space
$C^n$ of $n+1$-linear forms $\vp (x^0 , \ldots , x^n)$ on $A$, while
the basic operations are given by
\begin{equation}\label{ag}
\matrix{
(\delta_i  \vp) (x^0 , \ldots , x^n) &=& \vp (x^0 , \ldots , x^i
x^{i+1} , 
\ldots , x^n), \quad i=0,1,\ldots , n-1  \cr
\cr
(\delta_n  \vp) (x^0 , \ldots , x^n) &=& \vp (x^n  x^0 , x^1 , \ldots
, 
x^{n-1}) \hfill \cr
\cr
(\s_j  \vp) (x^0 , \ldots , x^n) &=& \vp (x^0 , \ldots , x^j , 1 ,
x^{j+1} 
, \ldots , x^n), \quad j=0,1,\ldots , n  \cr
\cr
(\tau_n  \vp) (x^0 , \ldots , x^n) &=& \vp (x^n , x^0 , \ldots ,
x^{n-1}) 
 . \hfill \cr
}
\end{equation}
\noindent These operations satisfy the relations (\ref{ad}) and (\ref{ae}). This
shows that any algebra $A$ gives rise canonically to a $\Lambda$-module and
allows \cite{C2,L} to interpret the cyclic cohomology groups $HC^n(A)$
as $Ext^n$ functors. All of the general properties of cyclic
cohomology such as the long exact sequence relating it to Hochschild
cohomology are shared by Ext of general $\Lambda$-modules and can be
attributed to the equality of the classifying space $B\Lambda$ of the
small category $\Lambda$ with the classifying space $BS^1$ of the
compact one-dimensional Lie group $S^1$.
One has
\begin{equation}\label{af}
B\Lambda = BS^1 =P_{\infty}(\Cb)
\end{equation}

\noindent For group rings $\Ac = \Cb \Gamma$ as above the cyclic cohomology 
bicomplex corresponds 
exactly (\cite{Bu}) to the bicomplex computing the $S^1$-equivariant cohomology of 
the free loop space
 of the classifying space $B\Gamma$, which is in essence dual to the space of 
irreducible representations of $\Gamma$.

\noindent In the recent years J. Cuntz and D. Quillen (\cite{CQ} \cite{C-Q$_2$} 
\cite{C-Q$_4$} ) have developed a powerful new approach 
to cyclic cohomology which allowed them to prove excision in full generality. A 
great deal of activity has also been generated around 
the work of Maxim Kontsevich on deformation theory and the Deligne conjecture on 
the 
fine structure of the algebra of Hochschild cochains (see \cite{Max}).
\bigskip

\section{Geometry}

\noindent The basic data of Riemannian geometry \cite{[R]}
consists of a manifold  $M$ whose points are locally labeled by a finite
number of real coordinates $\{x^{\mu}\}$ and  a {\it metric}, which is given by
the  infinitesimal line element:
\begin{equation}
ds^2 = g_{\mu \nu} \, dx^{\mu} \, dx^{\nu} \, . \label{eq:(1.1)}
\end{equation}
The distance between two points $x,y \in M$ is given by
\begin{equation}
d(x,y) = \hbox{Inf} \{ \text{Length} \ \g \, | \g \
\text{is a path between $x$ and $y$} \} \label{eq:(1.2)}
\end{equation}
where
\begin{equation}
\hbox{Length} \ \g = \int_\g ds \, . \label{eq:(1.3)}
\end{equation}
 One of the main virtues of Riemannian geometry is to be flexible enough
to give a good model of space-time in general relativity
(up to a sign change) while simple notions of Euclidean geometry continue to  make 
sense.
Homogeneous spaces which are geometries in the sense of the Klein program are too 
restrictive to achieve that goal.
For instance the idea of a straight line gives rise to the notion of geodesic and
the geodesic equation
\begin{equation}
{d^2 \, x^{\mu} \over dt^2} = -\G_{\nu \rho}^{\mu} \,
{dx^{\nu} \over dt} \ {dx^{\rho} \over dt}              \label{eq:(1.4)}
\end{equation}
where $\G_{\nu \rho}^{\mu} = {1\over 2} \, g^{\mu \a}
(g_{\a \nu ,\rho} + g_{\a \rho ,\nu} - g_{\nu \rho
,\a})$, gives the Newton equation of motion of a particle in
the Newtonian potential $V$ provided one uses the
metric $dx^2 + dy^2 + dz^2 - (1 +
2 V(x,y,z)) dt^2$ instead of the Minkowski metric (cf.\cite{[W]} for the more 
precise formulation).
\noindent  The next essential point is that the differential and integral  
calculus 
is available and
allows to go from the  local to the global.

\noindent The central notion of noncommutative geometry, comes from
the identification of the noncommutative analogue of the two basic concepts in 
Riemann's
formulation of Geometry, namely those of manifold and
of infinitesimal line element. Both of these noncommutative analogues are of 
spectral nature and combine 
to give rise to the notion of spectral triple and spectral manifold, which will be 
described in detail below.
\noindent  We shall first describe an operator theoretic framework for the 
calculus 
of infinitesimals which will 
provide a natural home for the line element $ds$.

\bigskip
\section{Calculus and Infinitesimals}

It was recognized at an early stage of the development of noncommutative geometry 
that the formalism of quantum mechanics gives a natural home both to 
infinitesimals
(the compact operators in Hilbert space) and to the integral (the logarithmic 
divergence in 
an operator trace) thus allowing for the  generalization of the differential and 
integral 
calculus which is vital for the development of the general theory. 

\smallskip

The following is the beginning of a long dictionary which translates
classical notions into the language of operators in the Hilbert space $\Hc$:
\[
\begin{array}{cc}
\hbox{Complex variable} &\hbox{Operator in} \
\Hc \\
 & \\
\hbox{Real variable} &\hbox{Selfadjoint operator} \\
 & \\
\hbox{Infinitesimal} &\hbox{Compact operator} \\
 & \\
\hbox{Infinitesimal of order} \ \a &\hbox{Compact operator with characteristic 
values} \\
&\mu_n \ \hbox{satisfying } \ \mu_n = O(n^{-\a}) \ , \
n\ra \ify \\
\hbox{Integral of an infinitesimal }
&{\int \!\!\!\!\! {{\scriptstyle -}}} \, T = \
\hbox{ Coefficient of logarithmic} \\
 \hbox{of order 1}
&\hbox{divergence  in the trace of } \ T \, . \\
\end{array}
\]

\smallskip

The first two lines of the dictionary are familiar from quantum
mechanics. The range  of a complex variable corresponds to the
{\it spectrum} of an operator. The holomorphic functional calculus gives a
meaning to $f(T)$ for all holomorphic functions $f$ on the spectrum of
$T$. It is only holomorphic functions which operate in this generality
which reflects the difference between  complex and real analysis.
When $T=T^*$ is selfadjoint then  $f(T)$ has a meaning for all
Borel functions $f$.

\noindent The size of the  infinitesimal $T \in \Kc$
is governed by the order of decay of the sequence of characteristic values 
$\mu_n = \mu_n (T)$  as $n \ra \ify$. In particular, for all
real positive $\a$ the following condition defines infinitesimals of order
$\a$:
\begin{equation}
\mu_n (T) = O (n^{-\a}) \qquad \hbox{when} \ n\ra
\ify                                    \label{eq:(2.10)}
\end{equation}
(i.e. there exists   $C>0$ such that  $\mu_n (T) \leq C
n^{-\a} \quad \fl \, n\geq 1$). Infinitesimals of order $\a$
also  form a two--sided ideal and moreover,
\begin{equation}
T_j \ \hbox{of order} \ \a_j \Ra T_1 T_2 \
\hbox{of order} \ \a_1 + \a_2 \, .              \label{eq:(2.11)}
\end{equation}

\smallskip

\noindent  Hence, apart from commutativity, intuitive
properties of the infinitesimal calculus are fulfilled.

\noindent  Since the size of an infinitesimal is measured by the
sequence $\mu_n \ra 0$ it might seem that one does not need the operator
formalism at all, and that it would be enough to replace the ideal $\Kc$ in
$\Lc (\Hc)$ by the ideal $c_0 (\Nb)$ of sequences converging to zero  in
the algebra $\ell^{\ify} (\Nb)$ of bounded sequences. A variable would just 
be a bounded sequence, and an infinitesimal a sequence $\mu_n, \ \mu_n
\downarrow 0$.  However, this commutative version does not allow for the
existence of variables with range a continuum since all elements of
$\ell^{\ify} (\Nb)$ have a point spectrum and a discrete spectral measure.
Only {\it noncommutativity} of $\Lc (\Hc)$ allows for the coexistence of variables 
with Lebesgue spectrum together with
infinitesimal variables. As we shall see shortly, it is precisely this lack of 
commutativity
between the line element and the coordinates on a space that will provide the 
measurement of 
distances.

\smallskip

\noindent The integral is obtained by the following analysis, mainly due to 
Dixmier 
(\cite{[Dx]}), of the logarithmic
divergence of the partial traces
\begin{equation}
\Trace_N (T) = \sum_{0}^{N-1} \mu_n (T) \ , \ T\geq 0 \,
.                                                       \label{eq:(2.22)}
\end{equation}
In fact, it is useful to define $\Trace_{\Lb} (T)$ for any positive real
$\Lb > 0$ by piecewise affine interpolation for noninteger $\Lb$. 
\smallskip

\noindent  Define for all order 1 operators   $T \geq 0$
\begin{equation}
\tau_{\Lb} (T) = {1\over \log \Lb} \, \int_e^{\Lb} \,
{\Trace_{\mu} (T) \over \log \mu} \ {d\mu \over \mu}
                                                        \label{eq:(2.26)}
\end{equation}
which is the Cesaro mean  of the function ${\Trace_{\mu} (T) \over \log
\mu}$  over  the scaling group $\Rb_+^*$.

\noindent For  $T \geq 0$, an infinitesimal of order 1, one has
\begin{equation}
\Trace_{\Lb} (T) \leq C \, \log \Lb                     \label{eq:(2.27)}
\end{equation}
so that $\tau_{\Lb} (T)$ is bounded.
The essential  property is the following  {\it
asymptotic additivity} of the coefficient $\tau_{\Lb} (T)$ of
the logarithmic divergence (\ref{eq:(2.27)}):
\begin{equation}
\vert \tau_{\Lb} (T_1 +T_2) - \tau_{\Lb} (T_1) -
\tau_{\Lb} (T_2) \vert \leq 3C \ {\log (\log \Lb) \over
\log \Lb}                                               \label{eq:(2.28)}
\end{equation}
for  $T_j \geq 0$.

\smallskip

An easy consequence of (\ref{eq:(2.28)}) is that any  limit point
$\tau$ of the nonlinear functionals $\tau_{\Lb}$ for ${\Lb}\ra \infty$
defines a positive and linear trace on the two--sided ideal of infinitesimals of 
order $1$,

\noindent In practice the choice of the limit point $\tau$ is irrelevant
because in all important examples $T$ is a {\it measurable }
operator, i.e.:
\begin{equation}
\tau_{\Lb} (T) \ \hbox{converges when } \ \Lb \ra
\ify \, .                                               \label{eq:(2.30)}
\end{equation}
Thus  the value $\tau (T)$ is independent of the choice of the limit point $\tau$ 
and is 
denoted
\begin{equation}
{\int \!\!\!\!\!\! -} \ T \, .                          \label{eq:(2.31)}
\end{equation}
 The first interesting example is provided by pseudodifferential
operators $T$ on a differentiable manifold $M$. When $T$ is of order 1
in the above sense, it is measurable and ${\int \!\!\!\!\!
-} T$ is the non-commutative residue of $T$ (\cite{[Wo]}). 
It has a local expression in terms of the distribution
kernel $k(x,y)$, $x,y \in M$. For $T$ of order $1$ the kernel $k(x,y)$
diverges logarithmically near the diagonal,
 \begin{equation}
k(x,y) = - a(x) \log \vert x-y \vert + 0(1) \ (\hbox{for} \ \  y \ra x)
                                                        \label{eq:(2.32)}
\end{equation}
 where   $ a(x)$ is a 1--density independent of the choice of Riemannian
distance $\vert x-y \vert$.
Then one has  (up to  normalization),
\begin{equation}
{\int \!\!\!\!\!\! -} \ T = \int_M a(x).                \label{eq:(2.33)}
\end{equation}
The right hand side of this formula makes sense for all
pseudodifferential operators (cf. \cite{[Wo]}) since one
can see that the kernel of such an operator is asymptotically of the form
\begin{equation}
k(x,y) = \sum a_k (x,x-y) - a(x) \log \vert x-y \vert +
0(1)                                                    \label{eq:(2.34)}
\end{equation}
where $a_k (x,\xi)$ is homogeneous of  degree $-k$ in $\xi$,
and the  1--density $a(x)$ is defined  intrinsically.

\smallskip

\noindent The same principle  of  extension of
${\int \!\!\!\!\! -}$ to infinitesimals of order
$<1$ works for hypoelliptic operators
and more generally as we shall see below, for  spectral triples
whose  dimension spectrum is simple.

\bigskip
\section{Manifolds}

 \noindent As we shall see shortly this framework gives a natural home for the 
analogue of 
the infinitesimal line element $ds$ of Riemannian geometry, but we need first to 
exhibit its 
compatibility with the 
notion of manifold.

\noindent It was recognized long ago
by geometors that the main quality of the
homotopy type of a manifold, (besides being
defined by a cooking recipee) is to satisfy
Poincar\'{e} duality not only in ordinary homology
but also in K-homology. 
Poincar\'e duality in ordinary homology is not sufficient to describe
homotopy type  of manifolds (\cite{Mi-S}) but D. Sullivan
(\cite{Sull}) showed (in the simply connected PL case of dimension
$\geq 5$ ignoring   2--torsion) that it is sufficient to replace ordinary
homology by $KO$--homology.

\noindent The characteristic property of {\it differentiable manifolds}
which is carried over to the  noncommutative case is
{\it Poincar\'e duality} in $KO$--homology.

\smallskip

Moreover, $K$-homology 
admits, as we saw above, a fairly simple definition in terms of Hilbert space 
Fredholm 
representations.

\noindent In the general framework of Noncommutative Geometry the confluence 
of the Hilbert space incarnation of the two notions
of  metric and fundamental class for a manifold led very naturally
to define a  geometric space as given by a {\it spectral triple:}
\begin{equation}
(\Ac ,\Hc ,D)                                           \label{eq:(1.6)}
\end{equation}
where $\Ac$ is an involutive algebra of operators in a Hilbert space
$\Hc$ and $D$ is a selfadjoint operator on $\Hc$. The involutive algebra
$\Ac$ corresponds to a given space $M$ like in the classical duality
``Space $\lra$ Algebra'' in algebraic geometry.  The infinitesimal line element in 
Riemannian 
geometry is given by the equality
\begin{equation}
ds=1/D,\label{E1}
\end{equation}
which expresses the infinitesimal line element $ds$ as the inverse of the Dirac
operator $D$, hence under suitable boundary conditions as a
propagator. 

\noindent The significance of $D$ is two-fold. On the one hand
it defines the metric by the above equation, on the other hand its
homotopy class represents the K-homology fundamental class of the
space under consideration.
The exact measurement of distances is performed as follows, 
instead of measuring distances between points using
the formula (5.2) we measure  distances between
states $\vp ,\psi$ on $\bar \Ac$ by a dual formula. This
dual formula involves {\it sup} instead of {\it inf} and does not use
paths in the space
\begin{equation}
d(\vp ,\psi) = \Sup \, \{ \vert \vp (a) - \psi (a) \vert \
; \ a\in \Ac \ , \ \Vert [D,a]\Vert \leq 1 \} \, .
                                                        \label{eq:(1.10)}
\end{equation}
A state, is a normalized positive  linear form on $\Ac$ such that $\vp (1) = 1$,
\begin{equation}
\vp : \bar{\Ac} \ra \Cb \ , \ \vp (a^* a) \geq 0 \ ,
\quad \fl \, a \in \bar{\Ac} \ , \ \vp (1) = 1 \, .
                                                        \label{eq:(1.9)}
\end{equation} In the commutative case the points of the space coincide with the 
characters of the algebra
 or equivalently with its pure states (i.e. the extreme points of the 
convex compact set of states).
As it should, this formula gives the geodesic distance
in the Riemannian case.  The spectral triple  $(\Ac ,\Hc ,D)$  associated 
to a compact  Riemannian manifold $M$,
$K$-oriented by a spin structure, is given by the representation
\begin{equation}
(f \, \xi) (x) = f(x) \, \xi (x) \qquad \fl \, x \in M \
, \ f\in \Ac \ , \ \xi \in \Hc                          \label{eq:(1.11)}
\end{equation}
of the algebra $\Ac$ of functions on  $M$ in the Hilbert space
\begin{equation}
\Hc = L^2 (M,S)                                         \label{eq:(1.12)}
\end{equation}
of square integrable sections of the spinor bundle.
The operator $D$ is the Dirac operator (cf. \cite{L-M}).
The commutator $[D,f]$, for $f\in \Ac= C^\infty(M)$
 is the Clifford multiplication by the gradient $\nb
f$ and its operator norm is:
\begin{equation}
\Vert [D,f]\Vert = \  \Sup_{x\in M}^{} \Vert \nb f (x)
\Vert = \hbox{Lipschitz  norm} \ f \, .                 \label{eq:(1.13)}
\end{equation}
Let $x,y \in M$ and $\vp ,\psi$ be the corresponding characters: $\vp (f) =
f(x)$, $\psi (f) = f(y) \ \text{for all}  \,  \ f \in \Ac.$ Then formula
(\ref{eq:(1.10)}) gives the same result as formula (5.2),
i.e. it gives  the geodesic distance between $x$ and $y$.

\smallskip

 \noindent Unlike the formula (5.2) the dual formula
(\ref{eq:(1.10)}) makes sense in general, namely, for example for discrete
spaces and even for totally disconnected spaces.

\smallskip

The second role of the operator $D$ is to define the fundamental class
of the space $X$ in K-homology, according to the following table,
\[
\begin{array}{cc}
\hbox{Space } \ X  & \hbox{Algebra} \ \Ac \\
& \\
K_1 (X)   &  \hbox{Stable homotopy class of the spectral} \\
&\qquad \hbox{triple} \ (\Ac ,\Hc ,D) \\
K_0 (X)  & \hbox{Stable homotopy class of } \Zb /2 \
\hbox{graded}\\
&  \hbox{ spectral triple}
\end{array}
\]
(i.e. for  $K_0$  we  suppose that $\Hc$ is  $\Zb /2$--graded by
 $\g$, where $\g = \g^*$, $\g^2 = 1$ and  $\g a
= a \g \quad \fl \,  a \in \Ac$, $\g D = -D\g$).

\smallskip

\noindent  This description works for the complex  $K$--homology
which is 2-periodic. We shall come back later to its refinement to $KO$-homology.

\bigskip
\section{Operator theoretic Index Formula}

 \noindent Before entering in the detailed discussion of the spectral notion of 
manifold
let us mention the local index formula. This result allows, using the 
infinitesimal 
calculus, 
to go from local to global in the general framework of spectral triples $(\Ac 
,\Hc,D)$. 

\smallskip

 \noindent The  Fredholm index of the operator  $D$
determines (in the odd case) an  additive map  $K_1 (\Ac) \
\stackrel{\vp}{\ra} \Zb$ given by the equality
\begin{equation}
\vp ([u]) = \Index \, (PuP) \ , \ u\in GL_1 (\Ac)
                                                \label{eq:(3.3)}
\end{equation}
where $P$ is the projector  $P = {1+F \over 2}$, $F = \Sign \, (D)$.

\smallskip
 
This map is computed by the pairing
of  $K_1 (\Ac)$ with the following  cyclic cocycle
\begin{equation}
\tau (a^0 ,\ldots ,a^n) = \Trace \, (a^0 [F,a^1] \ldots
[F,a^n]) \qquad \fl \, a^j \in \Ac
                                                \label{eq:(3.4)}
\end{equation}
where $F= \hbox{Sign} \ D$ and  we assume that the dimension $p$ of our space is 
finite,
which means that $(D+i)^{-1}$ is of order $1/p$, also $n\geq p$ is an odd integer. 
There are similar formulas involving the grading $\gamma$ in the even case, and it 
is quite satisfactory (\cite{theta} \cite{JLO}) that both cyclic cohomology and 
the 
chern Character formula adapt to the infinite dimensional case
 in which the only hypothesis is that $exp(-D^{2})$ is a trace class operator.

\smallskip

It is difficult to compute the cocycle $\tau$ in general because the formula
(\ref{eq:(3.4)}) involves the ordinary trace instead of the local trace
${\int \!\!\!\!\! -}$ and it is crucial to obtain a local form of the above 
cocycle.

\smallskip

 \noindent This problem is solved by a general formula \cite{[C-M2]} which we now 
describe

\noindent Let us make the following regularity hypothesis
on  $(\Ac ,\Hc ,D)$
\begin{equation}
a \ \hbox{and } \ [D,a] \ \in \ \cap \Dom \d^k , \ \fl \, a\in
\Ac                                                     \label{eq:(3.1)}
\end{equation}
where $\d$ is the derivation $\d(T) = [\vert D \vert,T]$ for any operator $T$.

\smallskip

\noindent We let $\Bc$ denote the algebra generated by  $\d^k (a)$,
$\d^k ([D,a])$. 
The usual notion of {\it dimension}  of a space is replaced by the
{\it dimension spectrum } which is a subset of $\Cb$.
\noindent The precise definition of the dimension spectrum is the subset
$\Si \sbs \Cb$ of singularities of the analytic functions
\begin{equation}
\z_b (z) = \Trace \, (b \vert D \vert^{-z}) \qquad \Re z
> p \ , \ b\in \Bc \, .
                                                        \label{eq:(3.2)}
\end{equation}
The dimension spectrum of a manifold $M$ is the set $\{
0,1,\ldots ,n\}$, $n=\dim M$; it is simple.  Multiplicities appear for
singular manifolds. Cantor sets provide examples of  complex points $z
\notin \Rb$ in the dimension  spectrum.

\noindent We assume  that $\Si$ is discrete and simple, i.e. that
$\z_b$ can be extended to  $\Cb / \Si$  with simple poles in  $\Si$.

\smallskip

\noindent We refer to  \cite{[C-M2]} for the case of a
spectrum with multiplicities.  Let $(\Ac ,\Hc ,D)$ be a spectral triple satisfying 
the hypothesis
(\ref{eq:(3.1)}) and (\ref{eq:(3.2)}). The local index theorem is the following,  
\cite{[C-M2]}:

\bigskip

\begin{enumerate}
\item The equality
\[
{\int \!\!\!\!\!\! -} P = \Res_{z=0} \, \Trace \, (P\vert
D\vert^{-z})
\]
defines a  trace on the algebra generated by
$\Ac$, $[D,\Ac]$ and $\vert D \vert^z$, where
$z\in \Cb$.

\item
There is  only  a finite number of non--zero terms in the following formula which
defines the odd components $(\vp_n)_{n=1,3,\ldots}$
of a cocycle in the bicomplex $(b,B)$ of $\Ac$,
 \[
\vp_n (a^0 ,\ldots ,a^n) = \sum_k c_{n,k}
{\int \!\!\!\!\!\! -} a^0 [D,a^1]^{(k_1)} \ldots
[D,a^n]^{(k_n)} \, \vert D \vert^{-n-2\vert k\vert}
\qquad \fl \, a^j \in \Ac
 \]
where the following notations are used:
$T^{(k)} = \nb^k (T)$ and $\nb (T) = D^2 T
- TD^2$, $k$ is a multi-index, $\vert k \vert = k_1 +\ldots + k_n$,
\[
c_{n,k} =
(-1)^{\vert k \vert} \, \sqrt{2i} (k_1 ! \ldots k_n
!)^{-1} \, ((k_1 +1) \ldots (k_1 + k_2 + \ldots + k_n
+n))^{-1} \, \G \left( \vert k \vert + {n\over 2}
\right).
\]

\item The pairing of the cyclic cohomology class $(\vp_n) \in HC^* (\Ac)$
with $K_1 (\Ac)$ gives the Fredholm index of $D$ with coefficients in
$K_1(\Ac)$.
\end{enumerate}

\bigskip

\noindent For the normalization of the pairing
between $HC^*$ and  $K(\Ac)$ see \cite{[Co]}.
In the even case, i.e. when $\Hc$ is $\Zb /2$ graded by $\g$,
 \[
\g = \g^*, \ \ \g^2 =1, \ \ \g a = a \g \quad \fl \, a \in \Ac, \ \g
D = -D\g,
\]
there is an analogous formula for a cocycle
$(\vp_n)$, $n$ even, which gives the Fredholm index of $D$
with  coefficients in  $K_0$. However,
$\vp_0$ is not expressed  in terms of the residue
${\int \!\!\!\!\! -}$ because it is not local for a
finite dimensional  $\Hc$.

\bigskip
\section{Diffeomorphism invariant Geometry}

The power of the above operator theoretic local trace formula lies in its 
generality.
We showed in  \cite{CM} how to use it to compute the index of transversally 
hypoelliptic
operators for foliations (\cite{[H-S]}). This allows to give a precise meaning to 
diffeomorphism invariant 
geometry on a manifold M, by the construction of a spectral triple $(\Ac ,\Hc ,D)$ 
where the 
algebra $\Ac$ is the crossed product of the algebra of smooth functions on the 
finite 
dimensional bundle $P$ of metrics on M by the natural action of the diffeomorphism 
group of 
M. The operator $D$ is an hypoelliptic operator which is directly associated to 
the 
reduction 
of the structure group of the manifold $P$ to a group of triangular matrices whose 
diagonal blocks are orthogonal. By construction the fiber of $P 
\stackrel{\pi}{\ra} 
M$ is the 
quotient $F/O(n)$ of the $GL(n)$--principal bundle $F$ of frames on $M$ by the 
action of the
orthogonal group $O(n) \sbs GL(n)$. The space $P$ admits a canonical foliation: 
the 
vertical 
foliation $V \sbs TP$, $V=\Ker \pi_*$ and   on the fibers $V$ and on  $N = (TP)/V$ 
the 
following Euclidean structures. A choice of $GL(n)$--invariant Riemannian metric
on  $GL(n)/O(n)$ determines a metric on  $V$. The metric on $N$ is defined 
tautologically: 
for every $p\in P$ one has a metric on  $T_{\pi (p)} (M)$ which is isomorphic to  
$N_p$  by 
$\pi_*$.

\noindent The computation of the local index formula for diffeomorphism invariant 
geometry 
\cite{CM} was quite complicated even in the case of codimension 1
foliations: there  were innumerable terms to be computed; this could be done by
hand, by 3 weeks of eight hours per day tedious computations, but it was of course 
hopeless to proceed by direct computations in the general case. Henri and I 
finally 
found how to get the answer for the general
case after discovering that the computation generated a Hopf algebra $\Hc(n)$ 
which 
only depends on n= codimension of
the foliation, and which allows to organize the computation provided cyclic 
cohomology is suitably adapted to Hopf algebras.

\noindent Hopf algebras arise very naturally from their actions on
noncommutative algebras \cite{[M]}. Given an algebra $A$, an 
action of the Hopf algebra $\Hc$ on $A$ is given by a linear map,
$$
\Hc \ot A \ra A, \quad h \ot a \ra h(a) 
$$
satisfying $h_1 (h_2  a) = (h_1  h_2) (a)$, $\fl  h_i \in {\Hc}$,
$a \in A$ and
\begin{equation}\label{ba}
h(ab) = \sum  h_{(1)}  (a)  h_{(2)}  (b)  \qquad \fl  a,b  \in A,
h \in {\Hc}.
\end{equation}
where the coproduct of $h$ is,
\begin{equation}\label{bb}
\D(h)=  \sum  h_{(1)}  \ot  h_{(2)} 
\end{equation}
In concrete examples, the algebra $A$ appears first, together with
linear maps $A \ra A$ satisfying a relation of the form (\ref{ba}) which dictates 
the Hopf algebra structure. This is exactly what occured in the above example (see 
\cite{CM} for the description of $\Hc(n)$ and its relation with Diff($\Rb^n$)).

\noindent The theory of characteristic classes for actions of $\Hc$
extends the construction \cite{C3} of cyclic cocycles from 
a Lie algebra of derivations of a $C^*$ algebra $A$, 
together with an \textit{invariant trace} $\tau$ on $A$.
 
\noindent This theory was developped in \cite{CM} in order to solve the above 
computational 
problem for diffeomorphism invariant geometry but it was shown in \cite{CM2} that 
the correct framework 
for the cyclic cohomology of Hopf algebras is that of modular pairs in involution. 
It is 
quite satisfactory that exactly the same structure
 emerged from the analysis of locally compact quantum groups.
The resulting cyclic cohomology appears to
be the natural candidate for the analogue of Lie algebra cohomology in
the context of Hopf algebras. We fix a group-like element $\sigma$ and a
character $\delta$ of $\Hc$ with $\delta(\sigma)=1$. They 
will play the role of the module of locally compact groups.

\noindent We then introduce the twisted antipode,
\begin{equation}\label{bc}
\wt S (y) = \sum  \delta (y_{(1)})  S (y_{(2)}) \ , \ y \in {\Hc}  , \
\D y = \sum  y_{(1)} \ot y_{(2)}.
\end{equation}

\noindent  We associate a cyclic complex (in fact a 
$\Lambda $-module, where $ \Lambda$ is the cyclic category), 
to any Hopf algebra together with a modular pair in involution. By this we mean a 
pair
($\sigma$, $\delta$) as above, such that the ($\sigma$, $\delta$)-twisted antipode 
is an involution,
\begin{equation}\label{ch} 
(\sigma^{-1}\wt{S})^2 = I.
\end{equation}
Then $\Hc_{(\delta,\sigma)}^{\natural} = \{ \Hc^{\ot n} \}_{n \geq 1}$ equipped 
with the operators given by the following formulas (\ref{ce})--(\ref{cg}) defines 
a 
module 
over the cyclic category $\Lambda$.
By transposing the standard simplicial operators underlying the 
Hochschild homology complex of an algebra,
one associates to $\Hc$, viewed only as a coalgebra, the 
natural cosimplicial  module $\{ \Hc^{\ot n} \}_{n \geq 1}$, with face operators 
$\delta_i: 
\Hc^{\ot n-1} \ra \Hc^{\ot n}$,
\begin{eqnarray}\label{ce}
&&\delta_0 (h^1 \ot \ldots \ot h^{n-1}) = 1 \ot h^1 
\ot \ldots \ot h^{n-1} \nonumber \\
&& \nonumber \\
&&\delta_j (h^1 \ot \ldots \ot h^{n-1}) = h^1 \ot \ldots \ot \D h^j \ot 
\ldots \ot h^n,\  \fl  1 \leq j \leq n-1, \nonumber\\
&&  \\
&&\delta_n (h^1 \ot \ldots \ot h^{n-1}) = h^1 \ot \ldots \ot h^{n-1}
\ot \sigma \nonumber
\end{eqnarray}
and degeneracy operators $\s_i : \Hc^{\ot n+1} \ra \Hc^{\ot n}$,
\begin{equation}\label{cf}
\s_i (h^1 \ot \ldots \ot h^{n+1}) = h^1 \ot \ldots \ot \ve (h^{i+1}) 
\ot \ldots \ot h^{n+1} \ , \ 0 \leq i \leq n.
\end{equation}

\noindent The remaining two essential features of a Hopf algebra 
--\textit{product} and \textit{antipode} -- are brought into play, to
define the \textit{cyclic operators} $\tau_n : \Hc^{\ot n} \ra \Hc^{\ot n}$,
\begin{equation}\label{cg} 
\tau_n (h^1 \ot \ldots \ot h^n) = (\D^{n-1}  \wt S (h^1)) \cdot h^2 \ot \ldots \ot 
h^n \ot 
\sigma.
\end{equation}
The theory of characteristic classes applies to actions of the Hopf algebra
on an algebra endowed with a $\delta$-invariant $\sigma$-trace.
A linear form $\tau$ on $A$ is a $\sigma$-trace
under the action of $\Hc$ iff one has,
$$
\tau (ab) = \tau (b \sigma (a)) \qquad \fl  a,b  \in A  .
$$
A $\sigma$-trace $\tau$ on $A$ is $\delta$-invariant
under the action of $\Hc$ iff 
$$
\tau (h(a)b) = \tau (a  \wt S (h)(b)) \qquad \fl  a,b  \in A  , \ h
\in 
{\Hc}. 
$$
The definition of the cyclic 
complex $HC^*_{(\delta,\sigma)} ({\Hc})$ is uniquely 
dictated in such a way that
the following defines a canonical map from $HC^*_{(\delta,\sigma)}
({\Hc})$ to 
$HC^* (A)$,
$$
\matrix{
\g (h^1 \ot \ldots \ot h^n) \in C^n (A)  , \ \g (h^1 \ot \ldots \ot
h^n) 
(x^0 , \ldots , x^n) = \cr
\cr
\tau (x^0  h^1 (x^1) \ldots h^n (x^n)). \cr
}
$$
\bigskip
\section{Hopf algebras, Renormalization and the 
Riemann-Hilbert problem}

I have been for many years fascinated by those topics in theoretical physics 
which combine mathematical sophistication together with validation by experiments. 
A prominent example is Quantum Field Theory, not in its abstract formulation but 
in its computational power, as a mysterious new calculus, known as perturbative 
renormalization.
It is heartening that
 some hard workers (\cite{Cc} \cite{rit}) continue to dig in the bottom of that 
mine 
and actually find gold.
 I had the luck to meet 
one of them, Dirk Kreimer, 
 and to join him in trying to unveil the secret beauty of these computations.

\noindent Dirk Kreimer showed (\cite{kkk} \cite{kk} \cite{k}) that for any quantum 
field theory, the
combinatorics of Feynman graphs is governed by a Hopf algebra $\Hc$ whose antipode 
involves the same 
algebraic operations as in the Bogoliubov-Parasiuk-Hepp recursion and the 
Zimmermann 
forest formula.

His Hopf algebra is commutative as an algebra and we showed in \cite{CK1} that it 
is 
the dual Hopf algebra of the
envelopping algebra of a Lie algebra $\underline G$ whose basis is labelled by
the one particle irreducible Feynman graphs. The Lie bracket of two such
graphs is computed from insertions of one graph in the other and vice
versa. The corresponding Lie group $G$ is the group of characters of
$\Hc$.

\smallskip

\noindent We also showed that, using dimensional regularization, the
bare (unrenormalized) theory gives rise to a loop

\begin{equation}
\g (z) \in G \ , \qquad z \in C
\end{equation}
where $C$ is a small circle of complex dimensions around the integer
dimension $D$ of space-time.
\noindent Our main result (\cite{ck} \cite{CK}) which relies on all the previous 
work of Dirk is that the renormalized theory is just the
evaluation at $z = D$ of the holomorphic part $\g_+$ of the Birkhoff
decomposition of $\g$.

\smallskip

\noindent The
Birkhoff decomposition is the factorization 
\begin{equation}
 \g \, (z) = \g_- (z)^{-1} \, \g_+ (z)
\qquad z \in C  
\end{equation}
 where we let $C \sbs P_1 (\Cb)$ be a
smooth simple curve, $C_-$ the component of the complement of $C$
containing $\ify \not\in C$ and $C_+$ the other component. Both
$\g$ and $\g_{\pm}$ are loops with values in $G$,
$$ \g \, (z) \in G  \qquad \fl \, z \in \Cb
 $$ and $\g_{\pm}$ are boundary values of holomorphic
maps (still denoted by the same symbol)
\begin{equation}
 \g_{\pm} : C_{\pm} \ra G \, .
\end{equation}
The normalization condition $\g_- (\ify) = 1$
ensures that, if it exists, the decomposition (2) is unique
(under suitable regularity conditions).
\noindent It is 
intimately tied up to the classification of holomorphic $G$-bundles
 on the Riemann sphere $P_1 (\Cb)$ and for $G= {\rm GL}_n (\Cb)$
to the Riemann-Hilbert problem.
\noindent  The Riemann-Hilbert problem comes
from Hilbert's $21^{\rm st}$ problem which he formulated as
follows:

\begin{itemize}
\item[] ``Prove that there always exists a Fuchsian linear
differential equation with given singularities and given monodromy.''
\end{itemize}

\noindent In this form it admits a positive answer due to Plemelj
and Birkhoff. When
formulated in terms of linear systems of the form, 
\begin{equation}
 y'(z) = A(z)
\, y(z) \ , \ A(z) = \sum_{\a \in S} \ \frac{A_{ \a} }{z - \a} \, ,
\end{equation}
 where $S$ is the given finite set of
singularities, $\ify \not\in S$, the $A_{\a}$ are complex matrices
such that 
\begin{equation}
 \sum \ A_{\a} = 0 
\end{equation} 
to avoid
singularities at $\ify$, the answer is not always positive (\cite {beauv}, \cite {Boli}), but the solution 
exists 
when the monodromy matrices
$M_{\a}$ are sufficiently close to 1. It can then
be explicitly written as a series of polylogarithms.

\noindent For $G = {\rm GL}_n (\Cb)$
the existence of the Birkhoff decomposition (2) is
equivalent to the vanishing,
\begin{equation}
c_1 \, (L_j) = 0 
\end{equation} 
of the Chern numbers $n_j = c_1 \, (L_j)$ of the holomorphic line
bundles of the Birkhoff-Grothendieck decomposition,
\begin{equation}
E = \op \, L_j 
\end{equation} 
where $E$ is the holomorphic vector bundle on $P_1 (\Cb)$ associated to
$\g$, i.e. with total space:
\begin{equation}
(C_+ \ts \Cb^n)\cup_{\g} (C_- \ts \Cb^n) \, .
\end{equation}

\smallskip

\noindent When $G$ is a simply connected nilpotent complex Lie group the
existence (and uniqueness) of the Birkhoff decomposition (2)
is valid for any $\g$. When the loop $\g : C \ra G$ extends to a
holomorphic loop: $C_+ \ra G$, the Birkhoff decomposition is given by
$\g_+ = \g$, $\g_- = 1$. In general, for $z \in C_+$ the evaluation,
\begin{equation}
\g \ra \g_+ (z) \in G 
\end{equation}
is a natural principle to extract a finite value from the singular
expression $\g (z)$. This extraction of finite values coincides with the
removal of the pole part when $G$ is the additive group $\Cb$ of complex
numbers and the loop $\g$ is meromorphic inside $C_+$ with $z$ as its
only singularity.

\noindent As I mentionned earlier our main result is that the renormalized theory 
is 
just the
evaluation at $z = D$ of the holomorphic part $\g_+$ of the Birkhoff
decomposition of the loop given by the unrenormalized theory $\g$.

\smallskip

\noindent We showed that the group $G$ is a
semi-direct product of an easily understood abelian group by a
highly non-trivial group closely tied up with groups of
diffeomorphisms, thanks to the relation that we had uncovered in \cite{CK1} 
between 
the Hopf algebra 
of rooted trees and the Hopf algebra $\Hc$ of section 9 involved in the 
computation 
of the index formula. 
The analysis of the relation between these two groups is intimately connected with the
 renormalization group and anomalous
dimensions, this will be the content of our coming paper \cite {cknew}.

\bigskip
\section{Spectral Manifolds}

\noindent
Let us now turn to manifolds and explain by giving concrete examples the content 
of 
our 
characterization (\cite{Co3}) of spectral triples associated to ordinary 
Riemannian 
manifolds.
It will be crucial that it applies to any Riemannian metric with fixed volume 
form. 
What we shall show in particular is that even in that classical case there is a 
definite advantage in 
dealing with the slightly noncommutative algebra of matrices of functions. The 
pair 
given by the algebra and the Dirac operator
 is then the solution of a remarkably simple polynomial equation. We shall also
 give a very natural "quantization" of the volume form of the manifolds which will 
appear
 most naturally in our examples, namely the spheres $S^n$ for n=1, 2 and 4.

\noindent Let us start with the simplest example, namely, let us show that the
geometry of the circle $S^1$ of length $2\pi$ is completely specified by the 
presentation:
\begin{equation}
U^{-1} [D,U] =1 \ , \ \hbox{where}\,  \ UU^* = U^* U =1 \, .
                                                        \label{eq:(4.1)}
\end{equation}
Of course $D$ is as above an unbounded selfadjoint operator.
We let $\Ac$ be the algebra of smooth functions
of the single element $U$. One has  $S^1 =
\hbox{Spectrum} \ (\Ac)$ as one easily checks using the invariance of the 
spectrum of $U$ by rotations implied by the above equation.
\noindent Any element $a$ of $\Ac$ is of the form $a = f(U)$ 
and one has
\begin{equation}
[D,a] = U \left( \frac{\partial}{\partial U} \, f \right) (U) = g \, (U) 
\label{eq12}
\end{equation}
and thus,
\begin{equation}
\Vert [D,a] \Vert = \ \build \Sup_X^{} \, \vert g(x) \vert \, , \quad g = U 
\, \frac{\partial}{\partial U} \, f \, . \label{eq13}
\end{equation}
This shows that the metric on $S^1 = \hbox{Spectrum} \ (\Ac)$ given by (6.3) is 
the 
standard Riemannian metric of length $2\pi$.
\noindent Let us now assume that  $ds = D^{-1}$ is an infinitesimal
of order 1. It is easy to see that this holds iff the commutant of the algebra 
generated by $U$ and $D$ is finite dimensional. We then claim that 
\begin{equation}
{\int \!\!\!\!\! {{\scriptstyle -}}} \, f \vert ds\vert = n \pi^{-1} \int \, f(x)  
\, \sqrt g 
\ d x \qquad \fl \, f \in \Ac  \label{eq:(98)}
\end{equation}
where the metric $g$ on $S^1$ is the above Riemannian metric and where the integer 
n 
is the 
index
\begin{equation}
n=- \Index \, (PUP) \ , 
                                                \label{eq:(99)}
\end{equation}
where $P$ is the projector  $P = {1+F \over 2}$, $F = \Sign \, (D)$.
This formula is simple to prove directly since it is enough to check it for 
irreducible pairs
$U, \, D$ in which case the spectrum of $D$ is of the form,
\begin{equation}
Spec(D)= \Zb + \lambda
                                                \label{eq:(199)}
\end{equation}
for some $\lambda $, while $U$ is the shift.

\noindent It is important for our later purpose to understand that it is a special 
case of 
the general index formula. Indeed both sides of  \ref{eq:(98)} are translation 
invariant and 
the equality for $f=1$ follows from
\begin{equation}
\Index \, (PUP) \, = -1/2 {\int \!\!\!\!\! {{\scriptstyle -}}} \, U^{-1} [D,U] \, 
\vert 
ds\vert   \label{eq:(200)}
\end{equation}
which follows from the following expression \cite{[Co]} for the n-dimensional 
Hochschild class of the 
Chern character of a spectral triple of dimension n,  
 \[
\tau_n (a^0 ,\ldots ,a^n) = 
{\int \!\!\!\!\!\! -}  a^0 [D,a^1] \ldots
[D,a^n] \, \vert D \vert^{-n}
\qquad \fl \, a^j \in \Ac
 \]
where we insert a $\gamma$ in the even case. This formula is weaker than the local 
index formula of section 8
 since it only gives the n-dimensional Hochschild class of the character, but it 
has 
the superiority to hold in full generality, 
with no assumption on the dimension spectrum. It is easy to use it to compute the 
index pairing with K-theory classes 
which come from the algebraic K-theory group $K_n$ since the Chern character of 
such 
classes is an n-dimensional Hochschild 
cycle. In the above toy example, $U$ defines an element in $K_1(\Ac)$ and its 
Chern 
character is the 1-dimensional Hochschild 
cycle $U^{-1} \ot U$ so that \ref{eq:(200)} follows.

\noindent Of course this toy example is a bit too simple, but the above K-theory 
discussion tells us how to proceed to higher dimension
 by relying on the formula (13) of section 4 for the Chern character and requiring 
the vanishing of the lower components. 

\noindent It is crucial that we do not restrict ourselves to the homogeneous case.

\noindent We shall now show that all geometries with fixed total area on the 
2-sphere $S^2$ are indeed
described by the following even analogues of equation (1),
\begin{equation}
\left\lgl e - \frac{1}{2} \right\rgl = 0 \, , \quad \left\lgl \left( e - 
\frac{1}{2} \right) \, [D,e] \, [D,e] \right\rgl = \g \label{eq24}
\end{equation}
where, as above, $D = D^*$ is an unbounded selfadjoint operator and $e$, $e^* 
= e$, $e^2 = e$ is a selfadjoint idempotent.

\noindent The right hand side of (\ref{eq24}) namely $\g$, is the $\Zb / 2$ 
grading of the Hilbert space $\Hc$ which is a characteristic feature of even 
dimensions, as we saw above. One has,
\begin{equation}
\g^2 = 1 \, , \ \g = \g^* \, , \ \g \, e = e \, \g \, , \ D \, \g = -\g \, D 
\, . \label{eq25}
\end{equation}
We still need to explain the symbol $\lgl T \rgl$ for any operator $T$ in 
$\Hc$. We fix a subalgebra $M \sbs \Lc (\Hc)$ isomorphic to $M_2 (\Cb)$ and let,
\begin{equation}
\lgl T \rgl = E_M \, (T) \label{eq26}
\end{equation}
where $E_M$ is the 
conditional expectation onto its commutant $M'$, given for instance as the 
integral over its unitary group of the conjugates $u \, T \, u^*$ of $T$. We 
assume that $D$ and $\gamma$ commute with $M$, 
\begin{equation}
D \in M' \, , \ \g \in M' \, . \label{eq27}
\end{equation}
One has the factorization $\Lc (\Hc) = M_2 (\Cb) \ot M'$, and any $T \in \Lc 
(\Hc)$ can be uniquely written as,
\begin{equation}
T = {\textstyle \sum} \ \ve_{ij} \, T^{ij} \qquad T^{ij} \in M' \label{eq28}
\end{equation}
where $\ve_{ij}$ are the usual matrix units in $M_2 (\Cb)$. We can apply 
(\ref{eq28}) to $T = e$ and we let $\Ac$ be the algebra of operators generated 
by the components $e^{ij}$ of $e$. Let us show that $\Ac$ is abelian and is 
the algebra of functions on the 2-sphere $S^2$.

\noindent We let $t = e^{11}$, $z = e^{12}$ so that
\begin{equation}
e^{22} = 1-t \, , \ e^{21} = z^* \label{eq29}
\end{equation}
using $\left \lgl e - \frac{1}{2} \right\rgl = 0$ and $e = e^*$. Also $t = 
t^*$ and $0 \, \build <_{{\textstyle =}}^{} \, t \, \build <_{{\textstyle 
=}}^{} \, 1$ follow from $e = e^*$ and $e^2 = e$. Thus $e = 
\left[ \matrix{t &z \cr z^* &(1-t) \cr} \right]$ and the equation $e^2 = e$ 
means that $t^2 + zz^* = t$, $tz + z \, (1-t) = z$, $z^* t + (1-t) \, z^* = z^*$,
$z^* z + (1-t)^2 = (1-t)$. This shows that $zz^* = z^* z$ and that $tz = zt$ 
so that $\Ac$ is abelian.

\noindent It also shows that the joint spectrum $X$ of $t$ and $z$ in $\Cb 
\ts \Cb$ is a compact subset of
\begin{equation}
\{ (t,z) \in [0,1] \ts \Cb \, ; \ (t^2 - t) + \vert z \vert^2 = 0 \} = P_1 
(\Cb) \, . \label{eq30}
\end{equation}
Let us now compute the left hand side of (\ref{eq24}) using $e = 
\left[\matrix{t &z \cr z^* &(1-t) \cr} \right]$ and the notation
\begin{equation}
dx = [D,x] \, . \label{eq31}
\end{equation}
We just expand the product of matrices,
\begin{equation}
\left[ \matrix{ \left( t - \frac{1}{2} \right) &z \cr z^* &\left( \frac{1}{2} 
- t \right) \cr} \right] \left[ \matrix{ dt &dz \cr dz^* &-dt \cr} \right] 
\left[ \matrix{ dt &dz \cr dz^* &-dt \cr} \right] \label{eq32}
\end{equation}
and take the sum of the diagonal elements. We get the terms,
\begin{eqnarray}
&&\left( t - \frac{1}{2} \right) (dt \, dt + dz \, dz^*) + z \, (dz^* \, dt - 
dt \, dz^*) \nonumber \\
&+ &z^* (dt \, dz - dz \, dt) + \left( \frac{1}{2} - t \right) (dz^* \, dz + 
dt \, dt) \nonumber \\
&= &\left( t - \frac{1}{2} \right) (dz \, dz^* - dz^* \, dz) + z \, (dz^* \, 
dt - dt \, dz^*) \nonumber \\
&+ &z^* (dt \, dz - dz \, dt) \, . \hfill \nonumber
\end{eqnarray}
Thus the second equation (\ref{eq24}) is equivalent to,
\begin{eqnarray}
&&\left( t - \frac{1}{2} \right) ([D,z] \, [D,z^*] - [D,z^*] \, [D,z]) + \\
&&z \, ( [D,z^*] \, [D,t] - [D,t] \, [D,z^*]) + \nonumber \\
&&z^* ([D,t] \, [D,z] - [D,z] \, [D,t]) = \g \, . \nonumber
\label{eq33}
\end{eqnarray}
Equivalently we can write it as,
\begin{equation}
\pi \, (c) = \g \label{eq34}
\end{equation}
where $c$ is the Hochschild 2-cycle,
\begin{equation}
c \in Z_2 (\Ac , \Ac) \label{eq35}
\end{equation}
given by the formula,
\begin{equation}
c = \left( t - \frac{1}{2} \right) \ot (z \ot z^* - z^* \ot z) + z \ot (z^* 
\ot t - t \ot z^*) + z^* \ot (t \ot z - z \ot t) \label{eq36}
\end{equation}
and where $\pi$ is the canonical map from Hochschild chains to operators 
(\cite{[Co]}),
\begin{equation}
\pi \, (a^0 \ot a^1 \ot \cdots \ot a^n) = a^0 \, [D,a^1] \cdots [D,a^n] 
\qquad \fl \, a^j \in \Ac \, . \label{eq37}
\end{equation}
We let $v \in C^{\ify} (S^2 , \wedge^2 \, T^*)$ be the 2-form on $S^2 = P_1 
(\Cb)$ associated to the Hochschild class of $c$ (\cite{Co$_{18}$}). It is given 
up 
to normalization by,
\begin{equation}
v = \frac{1}{1-2t} \ dz \wdg d \, \ov z \label{eq38}
\end{equation}
and vanishes nowhere on $S^2$.

\noindent We shall now show that {\it any}\/ Riemannian metric $g$ on $S^2$ 
whose associated volume form is equal to $v$,
\begin{equation}
\sqrt g \ d^2 x = v \label{eq39}
\end{equation}
gives canonically a solution to our equations (\ref{eq24})--(\ref{eq27}).

\noindent It is very important for our later considerations on gravity that 
not only the round metric but all possible metrics fulfilling (\ref{eq39}) 
actually appear as solutions.

\noindent The solution associated to a given metric $g$ fulfilling 
(\ref{eq39}) is constructed as follows, one lets
\begin{equation}
\Hc = L^2 (S^2 , S) \ot \Cb^2 \label{eq40}
\end{equation}
be the direct sum of two copies of the space of $L^2$ spinors on $S^2$. The 
algebra $M$ is just,
\begin{equation}
M = \Cb \ot M_2 (\Cb) \, . \label{eq41}
\end{equation}
The operator $D$ is given by,
\begin{equation}
D = \partial\!\!\!/ \ot 1 \label{eq42}
\end{equation}
where $\partial\!\!\!/$ is the Dirac operator (of the metric $g$). Finally the 
$\Zb / 2$ grading is
\begin{equation}
\g = \g_5 \ot 1 \label{eq43}
\end{equation}
where $\g_5$ is the chirality operator on spinors. 
 We identify $S^2$ with 
$P_1 (\Cb)$ which is the space
\begin{equation}
P_1 (\Cb) = \{ x \in M_2 (\Cb) \, , \ x^2 = x = x^* \, , \ \hbox{trace} \ x=1 
\} \label{eq44}
\end{equation}
and we let
\begin{equation}
e \in C^{\ify} (S^2) \ot M_2 (\Cb) \label{eq45}
\end{equation}
be the corresponding selfadjoint idempotent in $\Hc$ where $C^{\ify} (S^2)$ 
is acting by multiplication operators in $L^2 (S^2 , S)$.

\noindent One has (\ref{eq25})--(\ref{eq27}) by construction as well as 
$\left\lgl e - \frac{1}{2} \right\rgl = 0$ using (\ref{eq44}).

\noindent Let us check the second equality of (\ref{eq24}), or rather the 
equivalent form (17). For any $f \in \Ac = C^{\ify} (S^2)$ one has,
\begin{equation}
[D,f] = df \ot 1 \label{eq46}
\end{equation}
where $df = [\partial\!\!\!/ , f]$ is the Clifford multiplication by the 
differential of the function $f$.

\noindent For any $f^0 , f^1 , f^2 \in \Ac$ one has,
\begin{equation}
f^0 \, ([D,f^1] \, [D,f^2] - [D,f^2] \, [D,f^1]) = \rho \, \g \label{eq47}
\end{equation}
where $\rho$ is the smooth function such that
\begin{equation}
f^0 \, df^1 \wdg df^2 = \rho \, \sqrt g \ d^2 x \label{eq48}
\end{equation}
where $\sqrt g \ d^2 x$ is the volume form of the metric $g$. By (\ref{eq39}) 
we have $\sqrt g \ d^2 x = v$ and by construction of $v$ as the 2-form 
associated to the class of $C$ we get from (\ref{eq47}), (\ref{eq48}) that
\begin{equation}
\pi \, (c) = \rho \, \g \, , \qquad v = \rho \, v \label{eq49}
\end{equation}
i.e. $\rho = 1$.

\noindent This is enough to check that any Riemannian metric $g$ on $S^2$ 
with volume form equal to $v$ does give a solution of equations 
(\ref{eq24})--(\ref{eq27}).

\noindent To establish the converse one still needs technical assumptions in 
order to use the theorem of (\cite{Co3}), the main additional hypothesis being the 
order one condition which requires,
\begin{equation}
[[D,e^{ij}] , e^{k\ell}] = 0 \qquad \fl \, i,j,k,\ell \, . \label{eq50}
\end{equation}
Let us show now that the index formula (4) admits a perfect analogue 
in the general framework of solutions of (\ref{eq24})--(\ref{eq27}), assuming 
the following control of the dimension,
\begin{equation}
ds = D^{-1} \ \hbox{is of order} \ \frac{1}{2} \, , \label{eq51}
\end{equation}
i.e. the $n^{\rm th}$ characteristic value $\mu_n (D^{-1})$ is of order of 
$n^{-1/2}$ as $n \ra \ify$.

\noindent One has $e \in M_2 (\Ac)$ and the chern character of $e$ in the 
cyclic homology bicomplex $(b,B)$ is given by its components,
\begin{equation}
\left\lgl \left( e - \frac{1}{2} \right) \underbrace{e \ot \cdots \ot e}_{2n} 
\right\rgl = {\rm ch}_{2n} (e) \label{eq52}
\end{equation}
where the $\lgl \ \rgl$ means that we take the $M_2 (\Cb)$ trace of the 
corresponding elements.

\noindent Let us recall the index formula,
\begin{equation}
\hbox{Index} \ D_e^+ = \lgl {\rm ch} (e) , {\rm ch} (D) \rgl \label{eq53}
\end{equation}
which computes the index of the compression $e \, D^+ \, e$ of $D^+ : 
\frac{1+\g}{2} \, \Hc \ra \frac{1-\g}{2} \, \Hc$, in terms of the pairing 
between cyclic homology and cyclic cohomology. In general this requires the 
full knowledge of the chern character ${\rm ch} \, (D)$ in cyclic cohomology.

\noindent However in our case (\ref{eq24}) shows that ${\rm ch}_0 (e) = 0$, 
so that ${\rm ch}_2 (e)$ is a Hochschild cycle. Moreover by (\ref{eq51}) all 
the higher components of ${\rm ch} \, (D)$ vanish and (\cite{[Co]}) its 
component of degree 2, ${\rm ch}_2 (D)$ has a Hochschild class given by,
\begin{equation}
\tau_2 \, (a^0 , a^1 , a^2) = \, \int\!\!\!\!\!\!- \, \g \, a^0 \, [D,a^1] \, 
[D,a^2] \, D^{-2} \, . \label{eq54}
\end{equation}
The integral $\int\!\!\!\!\!-$ is a trace and when specializing (\ref{eq54}) 
to $a^j = e$ we can replace the integrand by its average $\left\lgl \left( e 
- \frac{1}{2} \right) \, [D,e] \, [D,e] \, D^{-2} \right\rgl = \g \, D^{-2}$.

\noindent Since $\g^2 = 1$ we thus obtain,
\begin{equation}
\int\!\!\!\!\!\!- \, ds^2 = \hbox{Index} \ D_e^+ \, . \label{eq55}
\end{equation}
In particular the area, taken in suitable units, is ``quantized'' by this 
equation since the index is always an integer.

\noindent This simple fact will take more meaning in the 4-dimensional case 
where the Einstein-Hilbert action will appear.

\noindent To close the discussion of this 2-dimensional example we note that 
the natural algebra here is not $\Ac$ but rather $M_2 (\Ac)$ which admits an 
amazingly simple presentation. It is generated by $M_2 (\Cb)$ and $e$ with 
the only relations
\begin{equation}
e = e^* \, , \ e^2 = e \, , \ \left\lgl e - \frac{1}{2} \right\rgl = 0 
\label{eq56}
\end{equation}
where $\lgl \ \rgl$ is the conditional expectation on the commutant of the 
subalgebra $M_2 (\Cb)$.

\noindent  Indeed the above computations show that the $C^*$ algebra 
generated by $M_2 (\Cb)$ and $e$ with the relations (\ref{eq56}) is,
\begin{equation}
C (S^2 , M_2 (\Cb)) = C (S^2) \ot M_2 (\Cb) \, . \label{eq57}
\end{equation}
Let us now move on to the 4-dimensional case.

\noindent We first determine the $C^*$ algebra generated by $M_4 (\Cb)$ and a 
projection $e = e^*$ such that $\left\lgl e - \frac{1}{2} \right\rgl = 0$ as 
above and whose matrix expression (\ref{eq28}) is of the form,
\begin{equation}
[ e^{ij} ] = \left[ \matrix{q_{11} &q_{12} \cr q_{21} &q_{22} \cr} \right] 
\label{eq58}
\end{equation}
where each $q_{ij}$ is a $2 \ts 2$ matrix of the form,
\begin{equation}
q = \left[ \matrix{\a &\b \cr -\b^* &\a^* \cr} \right] \, . \label{eq59}
\end{equation}
Since $e = e^*$, both $q_{11}$ and $q_{22}$ are selfadjoint, moreover since 
$\left\lgl e - \frac{1}{2} \right\rgl = 0$, we can 
find $t = t^*$ such that,
\begin{equation}
q_{11} = \left[ \matrix{t &0 \cr 0 &t \cr} \right] \, , \ q_{22} = \left[ 
\matrix{(1-t) &0 \cr 0 &(1-t) \cr} \right] \, . \label{eq60}
\end{equation}
We let $q_{12} = \left[ \matrix{\a &\b \cr -\b^* &\a^* \cr} \right]$, we then 
get from $e = e^*$,
\begin{equation}
q_{21} = \left[ \matrix{\a^* &-\b \cr \b^* &\a \cr} \right] \, . \label{eq61}
\end{equation}
We thus see that the commutant $\Ac$ of $M_4 (\Cb)$ is generated by $t,\a,\b$ 
and we need to find the relations imposed by the equality $e^2 = e$.

\noindent In terms of $e = \left[ \matrix{ t &q \cr q^* &1-t \cr} \right]$, 
the equation $e^2 = e$ means that $t^2 - t + q q^* = 0$, $t^2 - t + q^* q = 
0$ and $[t,q] = 0$. This shows that $t$ commutes with $\a$, $\b$, $\a^*$ and 
$\b^*$ and since $qq^* = q^* q$ is a diagonal matrix
\begin{equation}
\a \a^* = \a^* \a \, , \ \a \b = \b \a \, , \ \a^* \b = \b \a^* \, , \ \b 
\b^* = \b^* \b \label{eq62}
\end{equation}
so that the $C^*$ algebra $\Ac$ is abelian, with the only further relation, 
(besides $t = t^*$),
\begin{equation}
\a \a^* + \b \b^* + t^2 - t = 0 \, . \label{eq63}
\end{equation}
This is enough to check that,
\begin{equation}
\Ac = C (S^4) \label{eq64}
\end{equation}
where $S^4$ appears naturally as quaternionic projective space,
\begin{equation}
S^4 = P_1 (\Hb) \, . \label{eq65}
\end{equation}
The original $C^*$ algebra is thus,
\begin{equation}
B = C (S^4) \ot M_4 (\Cb) \, . \label{eq66}
\end{equation}
The analogue of (\ref{eq24}) is,
\begin{equation}
\left\lgl \left( e - \frac{1}{2} \right) \, [D,e]^{2n} \right\rgl = 0 \, , \ 
n = 0,1 \ \hbox{and} \ = \g \ \hbox{for} \ n=2 \, . \label{eq67}
\end{equation}
As above we assume,
\begin{equation}
D \in M' \, , \ \g \in M' \label{eq68}
\end{equation}
where $M = M_4 (\Cb)$ is the algebra of $4 \ts 4$ matrices. 

\noindent We shall first check by a direct computation that the equality 
$\left\lgl \left( e - \frac{1}{2} \right) \, [D,e]^2 \right\rgl = 0$ is 
automatic with our choice of $e$ (\ref{eq58}). We use (\ref{eq31}) for 
notational 
convenience and first compute exactly as in (\ref{eq32}), with $z$ replaced by 
$q = \left[ \matrix{ \a &\b \cr -\b^* &\a^* \cr} \right]$. We thus get,
\begin{eqnarray}
&\lgl (e - 1/2) \, [D,e]^2 \rgl &= \Biggl\lgl \left( t - \frac{1}{2} \right) 
\, (dq \, dq^* - dq^* \, dq) \\
&&+ \, q \, (dq^* \, dt - dt \, dq^*) + q^* \, (dt \, dq - dq \, dt ) 
\Biggl\rgl \nonumber
\label{eq69}
\end{eqnarray}
where the expectation in the right hand side is relative to $M_2 (\Cb)$.

\noindent The diagonal elements of $\om = dq \, dq^*$ are 
$$
\om_{11} = d \a \, d\a^* + d\b \, d\b^* \, , \ \om_{22} = d\b^* \, d\b + 
d\a^* \, d\a
$$
while for $\om' = dq^* \, dq$ we get, 
$$
\om'_{11} = d\a^* \, d\a + d\b \, d\b^* \, , \ \om'_{22} = d\b^* \, d\b + d\a 
\, d\a^* \, .
$$
It follows that, since $t$ is diagonal,
\begin{equation}
\left\lgl \left( t - \frac{1}{2} \right) \, (dq \, dq^* - dq^* \, dq) 
\right\rgl = 0 \, . \label{eq70}
\end{equation}
The diagonal elements of $q \, dq^* \, dt = \rho$ are
$$
\rho_{11} = \a \, d\a^* \, dt + \b \, d\b^* \, dt \, , \ \rho_{22} = \b^* \, 
d\b \, dt + \a^* \, d\a \, dt
$$
while for $\rho' = q^* \, dq \, dt$ they are
$$
\rho'_{11} = \a^* \, d\a \, dt + \b \, d\b^* \, dt \, , \ \rho'_{22} = \b^* \, 
d\b \, dt + \a \, d\a^* \, dt \, .
$$
Similarly for $\s = q \, dt \, dq^*$ and $\s' = q^* \, dt \, dq$ one gets the 
required cancellations so that,
\begin{equation}
\left\lgl \left( e - \frac{1}{2} \right) \, [D,e]^2 \right\rgl = 0 \, , 
\label{eq71}
\end{equation}
holds irrespective of the operator $D$ fulfilling (\ref{eq68}).

\noindent As in (\ref{eq38}) we let $v$ be the natural volume form on $S^4$ 
given by,
\begin{equation}
v = \frac{1}{1-2t} \ d\a \wdg d \, \ov{\a} \wdg d\b \wdg d \, \ov{\b} \, . 
\label{eq72}
\end{equation}
We shall now show that any Riemannian metric $g$ on $S^4$ whose associated 
volume form is $v$ gives a solution to (\ref{eq67}), (\ref{eq68}), thus,
\begin{equation}
\sqrt g \ d^4 x = v \, . \label{eq73}
\end{equation}
For this we proceed exactly as in (\ref{eq40})--(\ref{eq49}) above and we 
need to check that the Hochschild cycle $c$ obtained in the computation of 
\begin{equation}
\left\lgl \left( e - \frac{1}{2} \right) \, [D,e]^4 \right\rgl = \pi (c) 
\label{eq74}
\end{equation}
is totally antisymmetric, i.e. of the form,
\begin{equation}
c = \sum_{i,\s} \, \ve (\s) \, a_0^i \ot a_{\s (1)}^i \ot \cdots \ot a_{\s 
(4)}^i \label{eq75}
\end{equation}
where $\s$ ranges through all 24 permutations of $\{ 1 , \ldots , 4 \}$. With 
the above notations one has,
\begin{equation}
\left( e - \frac{1}{2} \right) \, [D,e]^4 = \left[ \matrix{ t - \frac{1}{2} 
&q \cr q^* &\frac{1}{2} - t \cr} \right] \, \left[ \matrix{ dt &dq \cr dq^* 
&-dt \cr} \right]^4 \label{eq76}
\end{equation}
and the sum of the diagonal elements is,
\begin{eqnarray}
&&\left( t - \frac{1}{2} \right) \, ((dt^2 + dq \, dq^*)^2 + (dt \, dq - dq 
\, dt) \, (dq^* \, dt - dt \, dq^*)) \nonumber \\
&- &\left( t - \frac{1}{2} \right) \, ((dt^2 + dq^* \, dq)^2 + (dq^* \, dt - 
dt \, dq^*) \, (dt \, dq - dq \, dt)) \nonumber \\
&+ &q \, ((dq^* \, dt - dt \, dq^*) \, (dt^2 + dq \, dq^*) + (dq^* \, dq + 
dt^2) \, (dq^* \, dt - dt \, dq^*)) \nonumber \\
&+ &q^* \, ((dt^2 + dq \, dq^*) \, (dt \, dq - dq \, dt) + (dt \, dq - dq \, 
dt) \, (dq^* \, dq + dt^2)) \, . \nonumber
\end{eqnarray}

Since $t$ and $dt$ are diagonal $2 \ts 2$ matrices of operators and the same 
diagonal terms appear in $dq \, dq^*$ and $dq^* \, dq$ as we saw in the proof 
of (\ref{eq70}), the first two lines only contribute by,
\begin{equation}
\left\lgl \left( t - \frac{1}{2} \right) \, (dq \, dq^* \, dq \, dq^* - dq^* 
\, dq \, dq^* \, dq ) \right\rgl \, . \label{eq77}
\end{equation}
Similarly the two last lines only contribute by,
\begin{eqnarray}
\label{eq78}
&&\lgl q^* \, (dt \, dq \, dq^* \, dq - dq \, dt \, dq^* \, dq + dq \, dq^* 
\, dt \, dq - dq \, dq^* \, dq \, dt) \\ 
&- &q \, (dt \, dq^* \, dq \, dq^* - dq^* \, dt \, dq \, dq^*
+ dq^* \, dq \, dt \, dq^* - dq^* \, dq \, dq^* \, dt ) \rgl \, . \nonumber
\end{eqnarray}
The direct computation of (\ref{eq77}) then gives
\begin{equation}
\sum \, \ve (\s) \left( t - \frac{1}{2} \right) \, d a_{\s (1)}^0 \, d a_{\s 
(2)}^0 \, d a_{\s (3)}^0 \, d a _{\s (4)}^0 \label{eq79}
\end{equation}
where $a_1^0 = \a$, $a_2^0 = \ov{\a}$, $a_3^0 = \b$, $a_4^0 = \ov{\b}$.

\noindent The direct computation of (78) gives
\begin{equation}
\sum_{i,\s} \, \ve (\s) \, a_0^i \, d a_{\s (1)}^i \, d a_{\s (2)}^i \, d 
a_{\s (3)}^i \, d a _{\s (4)}^i \label{eq80}
\end{equation}
where $i \in \{ 1,2,3,4 \}$ and,
\begin{eqnarray}
&&a_0^1 = \a \, , \ a_1^1 = t \, , \ a_2^1 = \ov{\a} \, , \ a_3^1 = \ov{\b} 
\, , \ a_4^1 = \b \nonumber \\
&&a_0^2 = \ov{\a} \, , \ a_1^2 = t \, , \ a_2^2 = \a \, , \ a_3^2 = \b \, , \ 
a_4^2 = \ov{\b} \nonumber \\
&&a_0^3 = \b \, , \ a_1^3 = t \, , \ a_2^3 = \ov{\b} \, , \ a_3^3 = \ov{\a} 
\, , \ a_4^3 = \a \nonumber \\
&&a_0^4 = \ov{\b} \, , \ a_1^4 = t \, , \ a_2^4 = \b \, , \ a_3^4 = \a \, , \ 
a_4^4 = \ov{\a} \, . \nonumber 
\end{eqnarray}
We thus obtain the required formula for the cycle $c$. When $dx = [D,x]$ with 
$D$ the Dirac operator associated to a Riemannian metric $g$ on $S^4$ we get 
as above, using the Clifford algebra, that
\begin{equation}
\pi (c) = \rho \, \g \label{eq81}
\end{equation}
where $\rho$ is the smooth function such that
\begin{equation}
\om = \rho \, \sqrt g \ d^4 x \label{eq82}
\end{equation}
where $\om$ is the differential form associated to $c$. Now, up to 
normalization one has,
\begin{eqnarray}
&\om &= \left( t - \frac{1}{2} \right) \, d\a \wdg d \, \ov{\a} \wdg d\b \wdg 
d \, \ov{\b} - \a \, dt \wdg d \, \ov{\a} \wdg d\b \wdg d \, \ov{\b} 
\nonumber \\
&&+ \, \ov{\a} \, dt \wdg d\a \wdg d\b \wdg d \, \ov{\b} - \b \, dt \wdg d \, 
\ov{\b} \wdg d \a \wdg d \, \ov{\a} \nonumber \\
&&+ \, \ov{\b} \, dt \wdg d\b \wdg d\a \wdg d \, \ov{\a} \, , \nonumber 
\end{eqnarray}
which using $t^2 - t + \a \, \ov{\a} + \b \, \ov{\b} = 0$ gives up to a factor 2,
\begin{equation}
\om =  \ \frac{1}{2t-1} \, d\a \wdg d \, \ov{\a} \wdg d\b \wdg d 
\, \ov{\b} \, . \label{eq83}
\end{equation}
Thus by hypothesis on $g$ we get $\rho = 1$ and $\pi (c) = \g$ which by the 
above computation means,
\begin{equation}
\left\lgl \left( e - \frac{1}{2} \right) \, [D,e]^4 \right\rgl = \g \, . 
\label{eq84}
\end{equation}

This shows that any Riemannian structure, with the given volume form on $M= S^4$, 
does give us a solution to 
our basic equation. Conversely exactly as in the two dimensional case we get, 
provided that $ds= D^{-1}$ is of order $\frac{1}{4}$, 
\begin{equation}
\int\!\!\!\!\!\!- \, ds^4 = - \hbox{Index} \ D_e^+ \, . \label{eq155}
\end{equation}
In particular the 4-dimensional volume, taken in suitable units, is ``quantized'' 
by 
this 
equation since the index is always an integer.

\noindent Let $\pi=(e, D, \gamma)$  be a solution of equations  (\ref{eq58}) 
(\ref{eq67}) (\ref{eq68}) above and let us assume  (\ref{eq50}), together
with harmless regularity conditions, \cite{Co3}.
Then there exists a unique Riemannian structure $g$ on $M$
such that the geodesic distance is given by
\[
d(x,y) = \Sup \, \{ \vert a(x) - a(y) \vert \ ; \ a\in
\Ac \ , \ \Vert [D,a] \Vert \leq 1 \} \, .
\]

\noindent The metric $g=g(\pi)$ depends only on the unitary
equivalence class of $\pi$. The fiber of the map
$\{$unitary equivalence
classes$\} \ra g(\pi)$ is an affine space
 $\Ac$ on which the functional  ${\int \!\!\!\!\! -} \,
ds^{2}$ is a positive quadratic form with  a unique real minimum  $\pi_{0}$ which 
is 
the representation described above
  in $L^2 (S^4,S)$ given by multiplication operators
and the Dirac operator associated to the Levi--Civita connection
of the metric $g$.

\noindent The value of  ${\int \!\!\!\!\! -} \,
ds^{2}$ on  $\pi_{o}$  is the Hilbert--Einstein action of the metric  $g$,
\[
{\int \!\!\!\!\!\! -} \, ds^{2} = -(48\pi^2)^{-1}\int r \,
\sqrt{g} \ d^4 x \ , .
\]
 We use the convention that the scalar curvature
$r$ is positive for the round sphere $S^4$, in particular,
the sign of the action
${\int \!\!\!\!\! -} \, ds^{2}$ is the correct one for
the  Euclidean formulation of gravity. We refer to \cite{Co3}, \cite{[K-W]}, 
\cite{[Kas]} for detailed computations.

\medskip

\bigskip
\section{Noncommutative Spectral Manifolds}

\noindent The main nuance in passing to the noncommutative case is that, since the 
diagonal in the square
 no longer corresponds to an algebra homomorphism (the map $ x \ot y \ra  \, xy$ 
is 
no longer an algebra homomorphism),
 the algebra $\Ac \ot \Ac^0$  now plays a central role.

\noindent The {\it fundamental class} of a noncommutative space (cf \cite{[Coo2]}),
is a class $\mu$ in the  $KR$--homology  of the algebra $\Ac \ot \Ac^0$
equipped with the  involution
\begin{equation}
\tau (x \ot y^0) = y^* \ot (x^*)^0 \qquad \fl \, x,y \in
\Ac                                                     \label{eq:(1.15)}
\end{equation}
where $\Ac^0$ denotes the algebra opposite to $\Ac$.  
The  $KR$-homology cycle representing $\mu$ is given by a spectral
triple, as above, equipped with an anti-linear
isometry $J$ on  $\Hc$ which implements the involution $\tau$,
\begin{equation}
J w J^{-1} = \tau (w) \qquad \fl \, w \in \Ac \ot \Ac^0
\, ,                                                    \label{eq:(1.16)}
\end{equation}
Instead of giving the action of the algebra $\Ac \ot \Ac^0$ in $\Hc$ one can 
equivalently
give an action of $\Ac$ satisfying the commutation rule,
$[a,b^0] = 0 \quad \fl \, a,b \in \Ac$ where
\begin{equation}
b^0 = J b^* J^{-1} \qquad \fl b \in \Ac
\end{equation}
$KR$-homology (\cite{[18]} \cite{[At]}) is periodic with period $8$ and the  
dimension modulo $8$ is specified by the
following commutation rules. One has $J^2 = \ve$, $JD = \ve' DJ$, $J\g = \ve'' \g 
J$
where  $\ve ,\ve' ,\ve'' \in \{ -1,1\}$ and with $n$ the dimension modulo 8,

\bigskip

\begin{center}
\begin{tabular}
{|c| r r r r r r r r|} \hline
{\bf n }&0 &1 &2 &3 &4 &5 &6 &7 \\ \hline
\hline
$\ve$  &1 & 1&-1&-1&-1&-1& 1&1 \\
$\ve'$ &1 &-1&1 &1 &1 &-1& 1&1 \\
$\ve''$&1 &{}&-1&{}&1 &{}&-1&{} \\  \hline
\end{tabular}
\end{center}

\bigskip

The anti-linear isometry $J$ is given in Riemannian geometry  by the
charge conjugation operator and in the noncommutative case by the Tomita-Takesaki 
antilinear 
conjugation operator \cite{T}. Given an involutive algebra of operators $\Ac$ on 
the 
Hilbert space $\Hc$, the Tomita-Takesaki theory associates to all  vectors  $\xi 
\in 
\Hc$,  \ 
cyclic for $\Ac$ and for its commutant $\Ac'$
\begin{equation}
\overline{\Ac \xi} = \Hc \ , \ \overline{\Ac' \xi} = \Hc
                                                        \label{eq:(4.4)}
\end{equation}
an anti-linear isometric involution $J:\Hc \ra
\Hc$ obtained from the polar decomposition of the operator
\begin{equation}
S \, a\xi = a^* \xi \qquad \fl \, a \in \Ac \ .
                                                        \label{eq:(4.5)}
\end{equation}
It satisfies the following commutation relation:
\begin{equation}
J \Ac'' J^{-1} = \Ac' \, .
                                                        \label{eq:(4.6)}
\end{equation}
In particular  $[a,b^0] = 0 \quad \fl \, a,b \in \Ac$ where
\begin{equation}
b^0 = J b^* J^{-1} \qquad \fl b \in \Ac
                                                        \label{eq:(4.7)}
\end{equation}
so $\Hc$ becomes an  $\Ac$-bimodule using the representation of
the opposite algebra. The class $\mu$ specifies only the stable homotopy class of 
the
spectral triple $(\Ac ,\Hc ,D)$ equipped with the isometry $J$ (and $\Zb
/2$--grading $\g$ if $n$ is even). The non-triviality of this homotopy
class shows up in the intersection form
\[
K_* (\Ac) \ts K_* (\Ac) \ra \Zb
\]
which is  obtained from the Fredholm index of $D$ with   coefficients
in $K_* (\Ac \ot \Ac^0)$. Note that it is  defined without using
the diagonal map $m:\Ac \ot \Ac \ra \Ac$, which is not a homomorphism
in the noncommutative case. This form is quadratic
or symplectic according to the value of $n$ modulo $8$.

 \smallskip

\noindent The  Kasparov intersection product \cite{[18]} allows to formulate
the Poincar\'e duality in terms of the invertibility  of $\mu$,
\[
\ex \, \b \in KR_n (\Ac^0 \ot \Ac) \ , \ \b \ot_{\Ac}
\mu = \id_{\Ac^0} \ , \ \mu \ot_{\Ac^0} \b = \id_{\Ac}
\, .
\]

\noindent It implies the isomorphism $K_* (\Ac) \
\stackrel{\cap \mu} \longra K^* (\Ac)$. 

\smallskip

\noindent  The condition that D is an operator of order one becomes
\[
[[D,a],b^0] = 0 \qquad \fl \, a,b \in \Ac \, .
                \]
(Notice that  since $a$ and  $b^0$ commute this condition
is equivalent to  $[[D,a^0],b]=0 \quad \fl \, a,b \in \Ac$.)

\smallskip

\noindent  One can show that the von Neumann algebra $\Ac''$ generated by $\Ac$ in
$\Hc$ is automatically finite and hyperfinite and there is a complete
list of such algebras up to isomorphism as we saw in section 2.  The algebra $\Ac$
is stable under smooth  functional calculus in its norm closure $A =
\bar{\Ac}$ so that $K_j (\Ac) \sm K_j (A)$, i.e. $K_j (\Ac)$ depends
only on the underlying topology (defined by the $C^*$ algebra $A$).
The integer $\chi = \lgl \mu ,\b \rgl \in \Zb$ gives the Euler characteristic in 
the 
form
\[
\chi = \Rang K_0 (\Ac) - \Rang K_1 (\Ac)
\]
and the general operator theoretic index formula of section 8 gives a local 
formula 
for $\chi$.

\smallskip

\noindent The group $\Aut^+ (\Ac)$ of automorphisms $\alpha$ 
of the involutive algebra $\Ac$, which are implemented by a unitary 
operator $U$ in $\Hc$ commuting with $J$,
\[
\alpha (x) = U \, x \, U^{-1} \qquad \fl \, x \in
\Ac \ , 
\]
plays the role of the group $\Diff^+ (M)$ of
diffeomorphisms preserving the K-homology fundamental class for a manifold $M$.

\smallskip

In the general noncommutative case, parallel to
the normal subgroup  $\Int \Ac \sbs \Aut \Ac$  of inner automorphisms of
$\Ac$,
\begin{equation}
\a (f) = ufu^* \qquad \fl \, f \in \Ac
                                                        \label{eq:(4.8)}
\end{equation}
where $u$ is a unitary element of $\Ac$ (i.e.
$uu^* = u^* u =1$), there exists a natural foliation
of the space of spectral geometries on
$\Ac$  by equivalence classes of
{\it inner deformations}
of a given geometry. To understand how they arise we need to understand 
how to transfer a given spectral geometry to a Morita
equivalent algebra. Given a spectral triple $(\Ac,\Hc ,D)$ and the Morita 
equivalence \cite{[Ri1]} between $\Ac$
and an algebra $\Bc$ where
\begin{equation}
\Bc = \End_{\Ac} (\Ec)
                                                        \label{eq:(4.16)}
\end{equation}
where  $\Ec$ is a finite,  projective, hermitian
right  $\Ac$--module, one gets a
spectral triple on  $\Bc$ by  the choice of  a {\it
hermitian connection} on  $\Ec$. Such a connection  $\nb$
is a linear map $\nb : \Ec \ra \Ec \ot_{\Ac} \Om_D^1$
satisfying the rules (\cite{[Co]})
\begin{equation}
\nb (\xi a) = (\nb \xi) a + \xi \ot da \qquad \fl \, \xi
\in \Ec \ , \ a\in \Ac
                                                        \label{eq:(4.17)}
\end{equation}
\begin{equation}
(\xi , \nb \eta) - (\nb \xi ,\eta) = d(\xi ,\eta) \qquad
\fl \, \xi ,\eta \in \Ec
                                                        \label{eq:(4.18)}
\end{equation}
where  $da = [D,a]$ and where $\Om_D^1 \sbs \Lc (\Hc)$ is
the $\Ac$--bimodule of operators of the form
\begin{equation}
A = \Si \, a_i [D,b_i] \ , \ a_i , b_i \in \Ac \, .
                                                        \label{eq:(4.10)}
\end{equation}

\smallskip

Any algebra $\Ac$ is Morita equivalent to itself (with $\Ec =
\Ac$) and  when one applies the above construction in the above context one gets 
the 
inner
deformations of the spectral geometry.

\noindent Such a deformation is obtained by
the following  formula (with suitable signs depending on the dimension mod 8) 
without modifying  neither the
representation of $\Ac$ in $\Hc$ nor the anti-linear isometry $J$
\begin{equation}
D\ra D+A+JAJ^{-1}
                                                        \label{eq:(4.9)}
\end{equation}
where $A=A^*$ is an arbitrary selfadjoint operator of the form \ref{eq:(4.10)}.
The action of the group
$\Int (\Ac)$ on the spectral geometries 
is simply the following gauge transformation of $A$
\begin{equation}
\g_u (A) = u[D,u^*] + uAu^* \, .
                                                        \label{eq:(4.11)}
\end{equation}
The required unitary equivalence is
implemented by the following representation of
the unitary group of
$\Ac$ in  $\Hc$,
\begin{equation}
u \ra uJuJ^{-1} = u(u^*)^0 \, .
                                                        \label{eq:(4.12)}
\end{equation}
The transformation (\ref{eq:(4.9)}) is the identity 
in the usual Riemannian case. To get a nontrivial example it suffices
to consider as we did in section 11, the product of a Riemannian triple by the 
unique spectral
geometry on the finite-dimensional algebra $\Ac_F = M_N (\Cb)$ of $N\ts N$
matrices on $\Cb$, $N\geq 2$. One then has $\Ac = C^{\ify} (M)
\ot \Ac_F$, $\Int (\Ac) = C^{\ify} (M,PSU(N))$ and inner deformations of
the geometry are parameterized by the gauge potentials for the gauge
theory of the group $SU(N)$.  The space of pure states of the algebra
$\Ac$, $P(\Ac)$, is the product $P = M\ts P_{N-1} (\Cb)$ and the metric
on $P(\Ac)$ determined by the formula (6.3) depends on the
gauge potential $A$. It coincide with the Carnot metric \cite{[G]} on $P$
defined by the horizontal distribution given by the connection
associated to $A$.  The group $\Aut (\Ac)$ of
automorphisms of $\Ac$ is the following semi--direct product
\begin{equation}
\Aut (\Ac) = \Uc \semi
\Diff^+ (M)
                                                        \label{eq:(4.13)}
\end{equation}
of the local gauge transformation group $\Int (\Ac)$ by the group of
diffeomorphisms. In dimension $n=4$, the Hilbert--Einstein action
functional for the Riemannian metric and the Yang--Mills action for
the  vector potential  $A$ appear with the correct signs in the
asymptotic expansion for large $\Lb$ of the number $N(\Lb)$ of
eigenvalues of $D$ which are $\leq \Lb$ (cf. \cite{[C-C]}), 
\begin{equation}
N(\Lb) = \# \ \hbox{eigenvalues of $D$ in} \ [-\Lb,\Lb] .
\end{equation}
This step function $N(\Lb)$ is the superposition of two terms,
$$
 N(\Lb)= \lgl N(\Lb) \rgl +N_{\rm osc} (\Lb).
$$
 The oscillatory part $N_{\rm osc} (\Lb)$ 
is the same as for a random matrix, governed by the statistic
dictated by the symmetries of the system and does not concern us here.
The average part $ \lgl N(\Lb) \rgl $ is computed by a semiclassical approximation 
from local expressions
 involving the familiar heat equation expansion. Other nonzero terms in
the asymptotic expansion  are  cosmological, Weyl gravity
and  topological terms. As we saw above in our characterization of section 11 we 
are 
only dealing with metrics with a fixed volume form so that the bothering 
cosmological term does not enter in the variational
equations associated to the spectral action $ \lgl N(\Lb) \rgl$. It is tempting to 
speculate that the phenomenological Lagrangian
 of physics, combining matter and gravity appears from the solution of an 
extremely 
simple 
operator theoretic equation along the lines described above.  As a starting point 
for such investigations see \cite{[Cncg]}.

\bigskip
\section{Noncommutative Tori}

A more sophisticated example of a spectral manifold is provided
by the noncommutative torus $\Tb_{\t}^2$.  The parameter $\t \in \Rb
/\Zb$ defines the following deformation of the algebra of smooth functions
 on the  torus  $\Tb^2$, with generators  $U,V$. The relations
\begin{equation}
VU=\exp 2\pi i \t \ UV \quad \hbox{and } \quad UU^* = U^*
U =1 \ , \ VV^* = V^* V =1
                                                        \label{eq:(4.15)}
\end{equation}
define the presentation \cite{C3} of the  involutive  algebra $\Ac_{\t} = \{ \Si 
\,
a_{n,m} U^n V^n \ ; \ a = (a_{n,m}) \in \Sc (\Zb^2)\}$ where $S(\Zb^2)$
is the Schwartz space of sequences with rapid decay.
We shall first describe a completely canonical procedure for 
constructing the $K$-cycle $(\hbox{\goth H} , D , \g)$ over ${\cal A}_{\t}$ 
from the fundamental class in cyclic cohomology, i.e., the choice of 
orientation, and the formal positive element
\begin{equation}
G = dU (dU)^* + dV (dV)^* \in \Om_+^2 ({\cal A}_{\t}) \, , \label{eq:(0.31)}
\end{equation}
which specifies the metric in the naive classical sense.

\noindent This transition from the $g_{\mu \nu}$ to the spectral triple extends in principle to 
arbitrary formal metrics $G \in \Om_+^2 ({\cal A}_{\t})$ but we stick to this specific flat example for simplicity.
The construction will be 
possible thanks to the noncommutative analogue of the Polyakov action of string 
theory.

\noindent We need first to explain briefly how this works in the commutative case.
The basic data is the fundamental class in cyclic cohomology, and the formal 
positive element
\begin{equation}
G = \sum_{\mu , \nu = 1}^{d}   g_{\mu \nu} dx^{\mu} (dx^{\nu})^* \in \Om_+^2 
({\cal 
A}) \, , \label{eq:(0.11)}
\end{equation}

\noindent  The first key notion is that of positivity in Hochschild cohomology.
 By definition (cf.~\cite{[Co-C]}) a Hochschild cocycle $\psi$ on a $*$-algebra 
${\cal A}$ is {\it positive} if it has {\it even} dimension $n = 2m$ and the 
following equality defines a positive sesquilinear form on the vector space 
${\cal A}^{\ot (m+1)}$:
\begin{equation}
\lgl a^0 \ot a^1 \ot \cdots \ot a^m , b^0 \ot b^1 \ot \cdots \ot b^m \rgl = 
\psi (b^{0*} a^0 , a^1 , \ldots , a^m , b^{m*} , \ldots , b^{1*}) 
\label{eq:(0.1)}
\end{equation}
for any $a^j , b^j \in {\cal A}$. 

\noindent  In general the positive Hochschild cocycles 
form a convex cone
\begin{equation}
Z_+^n ({\cal A} , {\cal A}^*) \sbs Z^n ({\cal A} , {\cal A}^*)
\label{eq:(0.4)}
\end{equation}
in the vector space $Z^n$ of Hochschild cocycles on ${\cal A}$.

\noindent Let $M$ be a 2-dimensional oriented compact 
manifold, ${\cal A}$ be the $*$-algebra of smooth functions on  $M$ 
and take for the class $C$ the fundamental class, i.e. the class of the de Rham 
current $C$
\begin{equation}
\lgl C , f^0 d f^1 \wedge df^2 \rgl = \frac{-1}{2\pi i} \int_M f^0 d f^1 \wedge 
df^2 \qquad \fl f^j \in C^{\ify} (M) \, . \label{eq:(0.7)}
\end{equation}
There is a natural correspondence between 
conformal structures on $M$ and {\it extreme points} of $Z_+^2 \cap C$. 
Thus, let $g$ be a conformal structure on $M$ or equivalently, since $M$ is 
oriented, a {\it complex} structure. Then, to the Lelong notion of positive 
current  corresponds the positivity in the above sense of the 
following Hochschild 2-cocycle:
\begin{equation}
\vp_g (f^0 , f^1 , f^2) = \frac{i}{\pi} \int_M f^0 \partial f^1 \wedge 
\ov{\partial} f^2 \, , \label{eq:(0.8)}
\end{equation}
where $\partial$ and $\ov \partial$ are inherited from the complex structure.
The mapping $g \mapsto \vp_g$ is an injection, since one can read off from $\vp_g$ 
what it 
means for a function $f$ to be holomorphic in a given small open set $U \sbs M$.
Each $\vp_g$ is an extreme point of the convex set $Z_+^2 \cap C$, and, 
conversely, the exposed points of this convex set can be determined as follows: 
for any element of the {\it dual cone} $(Z_+^2)^{\wedge}$ of $Z_+^2$, of the 
form
\begin{equation}
G = \sum_{\mu , \nu = 1}^{d} g_{\mu \nu} dx^{\mu} (dx^{\nu})^* \in \Om^2 ({\cal 
A}) \, , \label{eq:(0.11)}
\end{equation}
where $g_{\mu \nu}$ is a {\it positive} element of $M_d ({\cal A})$, one can 
show, assuming a suitable condition of nondegeneracy, that the linear form
\begin{equation}
\lgl G , \vp \rgl = \sum \vp (g_{\mu \nu} , x^{\mu} , (x^{\nu})^*) 
\label{eq:(0.12)}
\end{equation}
attains its minimum at a unique point in $Z_+^2 \cap C$, and that this point is 
equal to $\vp_g$, where $g$ is the conformal structure on $M$ associated with 
the classical Riemannian metric
\begin{equation}
g = \sum g_{\mu \nu} dx^{\mu} (dx^{\nu})^* \, . \label{eq:(0.13)}
\end{equation}
This allows us to understand the complex structures on $M$ as the solutions of 
a variational problem involving the fundamental class of $M$ and {\it 
positivity} in Hochschild cohomology. This problem is by no means restricted in 
its formulation to the {\it commutative} case, but it requires the notion of 
fundamental class in cyclic cohomology. It can be taken as a starting point for 
developing complex geometry in the noncommutative case.

\noindent Let us now show that the previous considerations extend without change 
to 
the 
noncommutative case and treat the noncommutative torus from a metric point of 
view.

\noindent The cyclic cohomology group $H C^0 ({\cal A}_{\t})$ is 1-dimensional and 
is 
generated by the unique trace $\tau_0$ of ${\cal A}_{\t}$,
\begin{equation}
\tau_0 \left( \sum a_{n,m} U^n V^m \right) = a_{0,0} \in \Cb \, , 
\label{eq:(0.20)}
\end{equation}
whereas the cyclic cohomology $HC^2 ({\cal 
A}_{\t})$ is two dimensional and besides $S \tau_0 \in HC^2$ (where $S$ is the 
periodicity operator in cyclic cohomology), is generated by the class of the 
cyclic 
2-cocycle
\begin{equation}
\tau_2 (a^0 , a^1 , a^2) = 2\pi i \sum_{n_0 + n_1 + n_2 = 0 \atop m_0 + m_1 + 
m_2 = 0} (n_1 m_2 - n_2 m_1) \, a_{n_0 , m_0}^0  a_{n_1 , m_1}^1  a_{n_2 , 
m_2}^2  \, . \label{eq:(0.21)}
\end{equation}
Note that only the {\it class} of this cocycle matters, not the above specific 
representative.
This nuance is very important since the above class only involves the smooth 
algebra 
${\cal A}_{\t}$; we 
shall now fix the metric. 
\begin{equation}
G = dU (dU)^* + dV (dV)^* \in \Om_+^2 ({\cal A}_{\t}) \, . \label{eq:(0.31)}
\end{equation}
On the intersection of the cyclic cohomology class $\tau_2 + b \, ({\rm Ker} B)$ 
with the positive cone $Z_+^2$ in Hochschild cohomology, the functional $G$ 
defined by
\begin{equation}
\vp \in Z^2 \mapsto \lgl G , \vp \rgl = \vp (1,U,U^*) + \vp (1,V,V^*) 
\label{eq:(0.23)}
\end{equation}
reaches its minimum at a unique point $\vp_2$ given by
\begin{equation}
\vp_2 (a^0 , a^1 , a^2) = 2\pi \sum_{n_0 + n_1 + n_2 = 0 \atop m_0 + m_1 + m_2 
= 0} (n_1 - im_1) (-n_2 - im_2) \, a_{n_0 , m_0}^0  a_{n_1 , m_1}^1  a_{n_2 , 
m_2}^2  \, . \label{eq:(0.24)}
\end{equation}

\noindent We then use the noncommutative analogue of a conformal structure, i.e., 
the positive cocycle $\vp_2$ together with the trace $\tau_0$, to construct the 
analogue of the Dirac operator for ${\cal A}_{\t}$, that is, we shall obtain a 
$(2,\ify)$-summable $K$-cycle $(\hbox{\goth H},D)$ on ${\cal A}_{\t}$. The 
Hilbert space $\hbox{\goth H}$ is the direct sum $\hbox{\goth H} = \hbox{\goth 
H}^+ \op \hbox{\goth H}^-$ of the Hilbert space $\hbox{\goth H}^- = L^2 ({\cal 
A}_{\t} , \tau_0)$ of the G.N.S. construction of $\tau_0$, and a Hilbert space 
$\hbox{\goth H}^+$ of forms of type $(1,0)$ on the noncommutative torus which is 
obtained 
canonically from $\vp_2$ as follows:
Let ${\cal A}$ be a $*$-algebra and let $\vp_2 \in Z_+^2 ({\cal A},{\cal A}^*)$ 
be a positive Hochschild $2$-cocycle on ${\cal A}$. Let $\hbox{\goth H}^+$ be 
the Hilbert space completion of $\Om^1 ({\cal A})$ equipped with the inner 
product
\begin{equation}
\lgl a^0 da^1 , b^0 db^1 \rgl = \vp_2 (b^{0*} a^0 , a^1 , b^{1*}) \, . 
\label{eq:(0.25)}
\end{equation}
Then the actions of ${\cal A}$ on $\hbox{\goth H}^+$ by left and right 
multiplications are unitary. 
They are automatically bounded if ${\cal A}$ is a pre-$C^*$-algebra.

\noindent Thus, $\hbox{\goth H}^+$ is a bimodule over ${\cal A}$ and the 
differential $d 
: {\cal A} \ra \Om^1 ({\cal A})$ gives a derivation which, for reasons that will 
become clear, we shall denote by $\partial : {\cal A} \ra \hbox{\goth H}^+$.

\noindent In our specific example, the computation is straightforward and gives 
$\hbox{\goth H}^+ = L^2 ({\cal A}_{\t} , \tau_0)$ as an ${\cal 
A}_{\t}$-bimodule and $\partial : {\cal A} \ra \hbox{\goth H}^+$ given by 
$\partial = \frac{1}{\sqrt{2\pi}} \, (\d_1 - i\d_2)$, where $\d_1$, $\d_2$ are 
the standard derivations of ${\cal A}_{\t}$.
\begin{equation}
\delta_1 = 2\pi i U \, \frac{\partial}{\partial U} \ , \delta_2 = 2\pi i V \,
\frac{\partial}{\partial V} \label{eq:(0.21)}
\end{equation}
so that $\delta_1 \left( \sum b_{nm} U^n V^m \right) = 2\pi i \sum n b_{nm}
U^n V^m$ and similarly for $\delta_2$. One has of course
\begin{equation}
\delta_1 \delta_2 = \delta_2 \delta_1 \label{eq:(0.22)}
\end{equation}
and the $\delta_j$ are derivations of the algebra ${\cal 
A}_{\t}$,
\begin{equation}
\delta_j (bb') = \delta_j (b) b' + b \delta_j (b') \quad \forall b , b' \in
{\cal A}_{\t} \, . \label{eq:(0.23)}
\end{equation} One can immediately check the 
following:
Let ${\cal A} = {\cal A}_{\t}$ act on the left on both $\hbox{\goth H}^- = L^2 
({\cal A}_{\t} , \tau_0)$ and $\hbox{\goth H}^+$. Then, the operator
\begin{equation}
D = \left[ \matrix{0 &\partial \cr \partial^* &0 \cr} \right] \label{eq:(0.26)}
\end{equation}
in $\hbox{\goth H} = \hbox{\goth H}^+ \op \hbox{\goth H}^-$ defines a 
$(2,\ify)$-summable $K$-cycle over ${\cal A}_{\t}$.
The $\Zb / 2$-grading $\g$ is given by the matrix $\g = \left[ \matrix{1 &0 \cr 
0 &-1 \cr} \right]$ and the real structure $J$ is given simply in terms of the 
Tomita-Takesaki antilinear isometry (cf. \cite{Co3}).

 \noindent  Translation invariant geometries on
$\Tb_{\t}^2$ are parameterized by complex numbers $\tau$ with positive
imaginary part like in the case of elliptic curves.  Up to isometry
the geometry depends only on the orbit of $\tau$ under the action of
$PSL (2,\Zb)$. However, a new phenomenon appears in the noncommutative case, 
namely,
the {\it Morita equivalence } which relates the  algebras $\Ac_{\t_1}$ and
$\Ac_{\t_2}$ if $\t_1$ and $\t_2$ are in the same orbit of the
$PSL (2,\Zb)$ action
on $\Rb$ \cite{[Ri1],[Ri2]}. We first need to give a concrete description of the 
finite projective modules over ${\cal A}_{\t}$, it is obtained by combining the 
results of \cite{C3} \cite{[26]} \cite{Rieffel}. 
\noindent The finite projective modules are classified up to isomorphism by a pair 
of integers $(p,q)$ such that $p+q \t \geq 0$.
Let us describe the simplest example of the modules ${\cal H}_{p,q}^{\t}$.
 The underlying linear
space is the usual Schwartz space,
\begin{equation}
{\cal S} (\Rb) = \{ \xi , \xi (s) \in \Cb \quad \forall s \in \Rb \}
\label{eq:(0.16)}
\end{equation}
of smooth function on the real line whose all derivatives are of rapid decay.

\noindent The right module structure is given by the action of the generators 
$U,V$
\begin{equation}
(\xi U) (s) = \xi (s+\theta) \ , \ (\xi V) (s) = e^{2\pi is} \xi (s) \quad
\forall s \in \Rb \, . \label{eq:(0.17)}
\end{equation}
One of course checks the relation (1), and it is a beautiful fact
that as a right module over ${\cal A}_{\t}$ the space \ref{eq:(0.16)} is {\it 
finitely
generated} and {\it projective} (i.e. complements to a free module). It
follows that it has the correct algebraic attributes to deserve the name of
``noncommutative vector bundle'' over $\Tb_{\theta}^2$ according to the first line 
of the dictionary of section 4,
$$
\matrix{
\hbox{Space $\, \Tb_{\theta}^2$} &\hbox{Algebra $\, {\cal A}_{\t}$} \cr
\cr
\hbox{Vector bundle} &\hbox{Finite projective module.} \cr
}
$$
The algebraic counterpart of a vector bundle $E$ on a space $X$ is its space of 
smooth sections
$C^{\ify} (X,E)$ and one can in particular compute its dimension by computing
the trace of the identity endomorphism of $E$. If one applies this method in
the above noncommutative example, one finds
\begin{equation}
{\rm dim}_{{\cal A}_{\t}} ({\cal S}) = \theta \, . \label{eq:(0.18)}
\end{equation}
The appearance of non integral dimension displays a
basic feature of von Neumann algebras of type II. The dimension of a vector
bundle is the only invariant that remains when one looks from the measure
theoretic point of view (Section 2). The von Neumann algebra which describes the 
noncommutative torus $\Tb_{\t}^2$
 from the measure theoretic point of view is the well known hyperfinite factor $R$ 
of type II$_1$. In particular the
classification of finite projective modules over $R$ is given by a positive
real number, the Murray and von Neumann {\it dimension},
\begin{equation}
{\rm dim}_R (\Ec) \in \Rb_+ \, . \label{eq:(0.20)}
\end{equation}
The next point (\cite{C3}) is that even though the {\it dimension} of the above 
module
is irrational, when we compute the analogue of the first Chern class, i.e. of
the integral of the curvature of the vector bundle, we obtain an integer.
We first need to determine the connections (in the sense of Section 12, equation 
10) 
on the finite projective module $\cal S$. It is not hard to see (using 17) that 
they 
are characterized
 by a pair of covariant differentials
\begin{equation}
\nabla_j : {\cal S} (\Rb) \rightarrow {\cal S} (\Rb) \label{eq:(0.24)}
\end{equation}
such that
\begin{equation}
\nabla_j (\xi b) = (\nabla_j \xi)b + \xi \delta_j (b) \quad \forall \xi \in
{\cal S} \ , b \in \Bc \, . \label{eq:(0.25)}
\end{equation}
One checks that, as in the usual case, the trace of the curvature $\Omega =
\nabla_1 \nabla_2 - \nabla_2 \nabla_1$, is independent of the choice of the
connection. Now the remarkable fact here is that (up to the correct powers of
$2\pi i$) the integral curvature of ${\cal S}$ is an integer. In fact for the
following choice of connection the curvature $\Omega$ is constant, equal to
$\frac{1}{\theta}$ so that the irrational number $\theta$ disappears in the
integral curvature, $\theta \times \frac{1}{\theta}$
\begin{equation}
(\nabla_1 \xi) (s) = - \frac{2\pi i s}{\theta} \, \xi (s) \ (\nabla_2 \xi)
(s) = \xi' (s) \, . \label{eq:(0.26)}
\end{equation}
Whith this integrality, one could get the wrong impression that the noncommutative 
torus $\Tb_{\theta}^2$
 looks very similar to the ordinary 2-torus. A striking difference is obtained by 
looking at the 
range of Morse functions. These are of course connected intervals for the
2-torus. For the above noncommutative torus the spectrum of a real valued
function such as
\begin{equation}
h = U + U^* + \mu (V+V^*) \label{eq:(0.27)}
\end{equation}
can be a Cantor set, i.e. have infinitely many disconnected pieces. This
shows that the one dimensional shadows of our space $\Tb_{\theta}^2$ are
considerably different from the commutative case. The above noncommutative
torus is the simplest example of noncommutative manifold, it arises naturally
not only from foliations but also from the Brillouin zone in the Quantum Hall
effect as understood by J. Bellissard, and in M-theory as we shall see in section 
14.

\noindent We shall now describe the natural moduli space (or more precisely, its 
covering 
Teichm\"uller space) for the noncommutative tori, together with a natural action 
of $SL(2,\Zb)$ on this space. The discussion parallels the description of the 
moduli space of elliptic curves but we shall find that our moduli space is the 
boundary of the latter space.

\noindent We first observe that as the parameter $\t \in \Rb / \Zb$ varies from 
$1$ 
to 
$0$ in the above labelling of finite projective modules ${\cal H}_{p,q}^{\t}$ one 
gets a monodromy, 
using the isomorphism $\Tb_{\t}^2 \sim \Tb_{\t + 1}^2$. The computation shows 
that this monodromy is given by the transformation $\left[ \matrix{ 1 &-1 \cr 0 
&1 \cr} \right]$ i.e., $x \ra x-y$, $y \ra y$ in terms of the $(x,y)$ 
coordinates in the $K$ group. This shows that in order to follow the 
$\t$-dependence of the $K$ group, we should consider the algebra ${\cal A}$ 
together with a choice of isomorphism,
\begin{equation}
K_0 ({\cal A}) \build \simeq_{}^{\rho} \Zb^2 \, , \quad \rho \ \hbox{(trivial 
module)} = (1,0) \, . \label{eq:(0.32)}
\end{equation}
Exactly as the Jacobian of an elliptic curve appears as a quotient of the 
$(1,0)$ part of the cohomology by the lattice of integral classes, we can 
associate canonically to ${\cal A}$ the following data:
\begin{enumerate}
\item[1)] The ordinary two dimensional torus $\Tb = HC_{\rm even} ({\cal A}) / 
K_0 ({\cal A})$ quotient of the cyclic homology of ${\cal A}$ by the image of 
$K$ theory under the Chern character map.
\item[2)] The foliation $F$ (of the above torus) given by the natural filtration 
of cyclic homology (dual to the filtration of $HC^{\rm even}$).
\item[3)] The transversal $T$ to the foliation given by the geodesic joining $0$ 
to the class $[1] \in K_0$ of the trivial bundle.
\end{enumerate}

It turns out that the algebra associated to the foliation $F$, and the 
transversal $T$ is isomorphic to ${\cal A}$, and that a purely geometric 
construction 
associates to every element $\a \in K_0$ its canonical representative from the 
transversal given by the geodesic joining $0$ to $\alpha$. (Elements of the 
algebra associated to the transversal $T$ are just matrices $a(i,j)$ where the 
indices $(i,j)$ are arbitrary pairs of elements $i,j$ of $T$ which belong to the 
same leaf. The algebraic rules are the same as for ordinary matrices. Elements 
of the module associated to another transversal $T'$ are rectangular matrices, and 
the dimension of the module is the transverse measure of $T'$.)

\noindent This gives the correct description of the modules ${\cal H}_{p,q}$. The 
above is in perfect analogy with the isomorphism of an elliptic curve with its 
Jacobian. The striking difference is that we use the {\it even} cohomology and 
$K$ group instead of the odd ones.

\noindent It shows that, using the isomorphism $\rho$, the whole situation is 
described by 
a foliation $dx = \t dy$ of $\Rb^2$ where the exact value of $\t$ (not only modulo 
$1$) does matter now.

\noindent Now the space of translation invariant foliations of $\Rb^2$ is the 
boundary $N$ 
of the space $M$ of translation invariant conformal structures on $\Rb^2$, and 
with $\Zb^2 \sbs \Rb^2$ a fixed lattice, they both inherit an action of $SL 
(2,\Zb)$. We now describe this action precisely in terms of the pair $({\cal A} 
, \rho)$. Let $g = \left[ \matrix{a &b \cr c &d \cr} \right] \in SL(2,\Zb)$. Let 
${\cal E} = {\cal H}_{p,q}$ where $(p,q) = \pm (d,-c)$, we define a new algebra 
${\cal A}'$ as the commutant of ${\cal A}$ in ${\cal E}$, i.e. as
\begin{equation}
{\cal A}' = {\rm End}_{{\cal A}} ({\cal E}) \, . \label{eq:(0.33)}
\end{equation}
It turns out (this follows from Morita equivalence) that there is a canonical map 
$\mu$ from $K_0 ({\cal A}')$ to $K_0 ({\cal A})$ (obtained as a tensor product 
over ${\cal A}'$) and the isomorphism $\rho' : K_0 ({\cal A}') \simeq \Zb^2$ is 
obtained by
\begin{equation}
\rho' = g \circ \rho \circ \mu \, . \label{eq:(0.34)}
\end{equation}
This gives an action of $SL(2,\Zb)$ on pairs $({\cal A} , \rho)$ with irrational 
$\t$ (the new value of $\t$ is $(a\t + b)/(c \t + d)$ and for rational values 
one has to add a point at $\ify$).

\noindent Finally another group $SL(2,\Zb)$ appears when we discuss the moduli 
space 
of 
flat metrics on $\Tb_{\t}^2$. Provided we imitate the usual construction of 
Teichm\"uller space by fixing an isomorphism,
\begin{equation}
\rho_1 : K_1 ({\cal A}) \ra \Zb^2 \label{eq:(0.35)}
\end{equation}
of the {\it odd} $K$ group with $\Zb^2$, the usual discussion goes through and 
the results of \cite{Co3} show that for all values of $\t$ one has a canonical 
isomorphism of the moduli space with the upper half plane $M$ divided by the 
usual action of $SL(2,\Zb)$. Moreover, one shows that the two actions of 
$SL(2,\Zb)$ 
actually commute. The striking fact is that the relation between the two 
Teichm\"uller spaces,
\begin{equation}
N = \partial M \label{eq:(0.36)}
\end{equation}
is preserved by the diagonal action of $SL (2,\Zb)$.
Finally note that the above action of $SL (2,\Zb)$ on the parameter $\t$ lies beyond the 
purely formal realm of deformation theory in which $\t$ is treated as a formal deformation parameter.
This is a key point in which noncommutative geometry should be distinguished from formal atempts 
to deform standard geometry. 
\bigskip
\section{Noncommutative gauge Theory and String Theory}

The analogue of the Yang-Mills action functional and the classification of 
Yang-Mills connections on the 
noncommutative tori was developped in \cite {Co-R}, with the primary goal of 
finding 
a "manifold shadow" for these noncommutative spaces.
These moduli spaces turned out indeed to fit this purpose perfectly, allowing for 
instance to find the usual Riemannian space of gauge equivalence classes of 
Yang-Mills connections as an invariant of the noncommutative metric. The next 
surprise came from the natural occurence (as an unexpected guest) of 
both the noncommutative tori and the components of the Yang-Mills connections in 
the 
classification of the BPS states in M-theory \cite{CDS}.
 In the matrix formulation of M-theory the basic equations to obtain periodicity 
of 
two of the basic coordinates $X_i$ turns out to be the 
following variant of equation 1 of section 11, 
\begin{equation}
U_i X_j U_i^{-1} =X_j + a \delta_i^j , i= 1,2  \label{eq:(0.136)}
\end{equation}
where the $U_i$ are unitary gauge transformations. 

\noindent The multiplicative commutator $U_1 U_2 U_1^{-1}U_2^{-1}$ is then central
and in the irreducible case its scalar value $ \lambda =\exp 2\pi i \t $ brings in 
the algebra of coordinates on the noncommutative torus.
The $X_j$ are then the components of the Yang-Mills connections. It is quite 
remarkable that the same picture emerged from the
other information one has about M-theory concerning its relation with 11 
dimensional 
supergravity and that string theory dualities could be interpreted
using Morita equivalence. The latter relates as we saw above in section 13 the 
values of $\theta$ on an orbit of $SL (2,\Zb)$ and this
 type of relation would be invisible in a purely deformation theoretic 
perturbative 
expansion like the one given by the Moyal product.

\smallskip

\noindent In their remarkable paper, Nekrasov and Schwarz \cite{N-S} showed that 
Yang-Mills gauge theory on
 noncommutative $\Rb^4$ gives a conceptual understanding of the nonzero B-field 
desingularization of the 
moduli space of instantons obtained by perturbing the ADHM equations.
In their paper \cite{Witten}, Seiberg and Witten exhibited the unexpected relation 
between the 
standard gauge theory and the noncommutative one.

\noindent   The question of renormalizability of quantum field theories on 
noncommutative spaces \cite{Filk} \cite{Kraj} \cite{Gros} \cite{Chepelev} 
\cite{Gracia} which was the basis of \cite{Co$_{28}$} is generating 
remarkable similarities with string theory \cite{Seiberg} which 
hopefully should yield a better formulation of $M$-theory than what is 
currently available. 
The rate at which progress is occuring in this interplay between 
noncommutative geometry and physics makes it rather futile to try and foresee 
what will happen even in the near future but there are a few issues on which I cant help to make  
brief comments (as a non-expert). The first has to do with locality, the expressions discussed in 
section 8 which 
involve the residue applied to multiple products of elements of the algebra and 
the 
operator $D$ do
 generate the natural candidate for local cochains in the general case. This was 
the 
basic procedure used in \cite{Co$_{28}$} to generate {\it local} interactions.

\noindent Also the transformation from one standard gauge theory to the 
noncommutative one in \cite{Witten} has the 
basic feature of respecting the foliations of gauge potentials by gauge 
equivalence 
and since gauge transformations are isospectral deformations 
of the corresponding Dirac operators (with potential) it is natural to wonder 
wether 
the Seiberg-Witten transformation 
can be interpreted in spectral terms.

\noindent String theory is a generalization of 
ordinary geometry whose onshell formulation is understood via conformal field 
theory. The corresponding mathematical question of existence 
 of $\sigma $-models should benefit from the investigation of the Riemann-Hilbert 
problem attached to the renormalization of such a theory as in section 10. 

\noindent
Finally, one should probably also 
 look for an offshell formulation of string-geometry. It is well known that the 
spectral information on a homogeneous Riemannian
space can be grasped using Lie group representations but what we showed in section 
11 is that even the nonhomogeneous metrics
are accessible to such a Hilbert space representation treatment. The new feature is that the basic 
equations are no longer related to Lie group representations but to
algebraic K-theory considerations.
\noindent It is tempting to speculate that a similar adaptation of the Lie algebra 
representation theoretic approach to conformal field theory could
yield the desired offshell formulation of stringy geometry.
\vglue 1cm



\end{document}


\bibitem{[D-F-R]} S. Doplicher, K. Fredenhagen and J.E.
Roberts: Quantum structure of space time at the Planck
scale and Quantum fields, to appear in CMP.

\bibitem{[G-F]} K. Gawedzki and J. Fr\"ohlich: Conformal
Field theory and Geometry of Strings, {\it CRM Proceedings
and Lecture Notes}, Vol.7 (1994), 57-97.f

\bibitem{AS} Schwarz, A., Morita equivalence and duality,  
{\it Nucl. Phys.} {\bf B534} (1998), 720-738; hep-th/9805034.